\documentclass[11pt]{article}

\usepackage{amssymb,latexsym, amsmath}
\usepackage{graphicx}

\makeatletter
\@addtoreset{figure}{section}
\def\thefigure{\thesection.\@arabic\c@figure} \def\fps@figure{h, t}
\@addtoreset{table}{bsection}
\def\thetable{\thesection.\@arabic\c@table} \def\fps@table{h, t}
\@addtoreset{equation}{section}

\makeatother

\advance \topmargin by -\headheight
\advance \topmargin by -\headsep
\evensidemargin \oddsidemargin
\newfont{\tenbi}{cmbxti10}

\begin{document}

\title{On Soliton-type Solutions of Equations Associated with N-component
  Systems \footnote{PACS numbers 03.40.Gc, 11.10.Ef, 68.10.-m, AMS
    Subject Classification 58F07, 70H99, 76B15}}

\author {Mark S. Alber \thanks{Research partially supported by NSF grant DMS
    9626672.} \\ Department of Mathematics\\ University of Notre Dame\\
  Notre Dame, IN 46556 \\ {\footnotesize Mark.S.Alber.1@nd.edu} \and Gregory
  G. Luther
  \thanks{Research partially supported by NSF grant DMS 9626672.} \\
  Engineering Sciences and Applied Mathematics Department\\ McCormick School
  of Engineering and
  Applied Science\\
  Northwestern University\\ 2145 Sheridan Road\\ Evanston, Il 60208--3125\\
  {\footnotesize ggluther@nwu.edu} \and Charles Miller\\ Department of
  Mathematics\\ University of Notre Dame\\ Notre Dame, IN 46556 \\
  {\footnotesize cmiller6@nd.edu} \\}

\maketitle

\begin{center} {byline: Soliton-type Solutions and $N$-component
  Systems} \end{center}
\begin{abstract} The algebraic geometric approach to $N$-component systems
  of nonlinear integrable PDE's is used to obtain and analyze explicit
  solutions of the coupled KdV and Dym equations.  Detailed analysis of
  soliton fission, kink to anti-kink transitions and multi-peaked soliton
  solutions is carried out.  Transformations are used to connect these
  solutions to several other equations that model physical phenomena in
  fluid dynamics and nonlinear optics.
\end{abstract}

\section{Introduction}
The solution of nonlinear evolution equations using techniques from
algebraic geometry was initially developed to handle $N$-phase wave trains.
With this approach, parameterized families of quasi-periodic and soliton
solutions are associated with Hamiltonian flows on level sets of
finite-dimensional phase spaces. In Section \ref{sec:one}, these flows are
described using the $\mu$ variable representation on symmetric products of
Riemann surfaces. The first integrals in the quasi-periodic case have the
form $P_j^2 = C(\mu_j)$ where $C(\mu_j)$ is a polynomial with constant
coefficients and $P_j$ is the conjugate variable for $\mu_j$.  The
polynomial $C(E)$ is called the spectral polynomial and determines the
form of the first integrals.

The algebraic geometric approach provides a way to construct solutions by
analytical or numerical integration of a system of Hamiltonian equations for
these $\mu$ variables. Solutions of the nonlinear PDE's are expressed in
terms of these variables by using the trace formulas.  The $\mu$ variable
representation yields an action angle representation on a Jacobi variety
(invariant variety in the phase space) that linearizes the Hamiltonian flow,
and in the quasi-periodic case the solution of the $\mu$ equations is
reduced to a Jacobi inversion problem. Solutions are then expressed using
Riemann theta functions. For details see, for example, Mumford(1983) and
Ercolani and McKean (1990).

As pairs of roots of the polynomial $C(E)$ coalesce, soliton solutions begin
to appear.  Applying this limit to the first integrals and the equations of
motion in terms of the $\mu$ variables for quasi-periodic solutions,
Hamiltonian systems of ODE's that describe soliton solutions are obtained.
Soliton solutions are computed by solving these equations and using the
trace formula to connect them to the associated nonlinear PDE's.  In the
soliton limit the Jacobi inversion problem for the system often reduces to a
system of algebraic equations.  As will be shown, these algebraic equations
are exactly solvable in the case of genus one and two.  For details about
the connection between soliton and quasi-periodic solutions see for example
Ablowitz and Ma (1981) and Alber and Alber(1985).  The exact representations
obtained for soliton solutions are shown to be related to the Hirota
$\tau$-functions.  The case of genus $n$ solutions is analogous to the lower
genus case and can be described using similar formulas.  Using the
action-angle representation, the algebraic geometric approach also
introduces a powerful way to compute the phase shifts due to soliton
interactions (see Alber and Marsden(1992) for details.)

In this paper, soliton solutions of multicomponent systems of equations are
studied using the algebraic geometric approach. The soliton fission effect,
kink to anti-kink transitions, and multi-peaked solitons are demonstrated
using a class of commuting Hamiltonian systems on Riemann surfaces. These
first two effects manifest themselves in the soliton limit of the genus two
quasi-periodic solution when six roots of the spectral polynomial $C(E)$
coalesce in a pairwise fashion.  Exact formulas for these solutions are
obtained and asymptotic and numerical analysis of them is performed.  The
technique used to obtain these limiting solutions is demonstrated explicitly
in the case of the coupled Korteweg-de~Vries (cKdV) and the coupled Dym
(cDym) equations and physically relevant equations associated with them.

The modified cKdV system is
\begin{eqnarray} \label{cKdVinit}
  u_t &=& v_x -{\textstyle \frac{3}{2}} uu_x + K_1 u_x\  ,\\
  v_t &=& {\textstyle\frac{1}{4}}u_{xxx} -vu_x
  -{\textstyle\frac{1}{2}}uv_x +K_1 v_x \ ,
\end{eqnarray}
which reduces to the cKdV system for $K_1=0$. Otherwise $K_1$ is viewed as a
small constant. The coupled Dym equations are given by
\begin{eqnarray}
  u_t&=&{\textstyle
  \frac{1}{4}}u_{xxx}-{\textstyle\frac{3}{2}}uu_x+v_x+K_1u_x \
  ,\\
  v_t&=&-u_xv-{\textstyle\frac{1}{2}}uv_x+K_1v_x\ .
\end{eqnarray}
The general method is demonstrated by describing quasi-periodic and soliton
solutions for the cKdV and cDym systems for genus two and less.

The cKdV and cDym equations are both generic examples of $N$-component
systems. Energy dependent Schr\H{o}dinger operators and $bi$-Hamiltonian
structures for multicomponent systems were investigated in Antonowicz and
Fordy(1987).  Quasi-periodic and soliton solutions were studied in
connection with Hamiltonian systems on Riemann surfaces in Alber {\it
  et~al.}(1997).  In Alber {\it et al.}  (1994) it was also shown that the
presence of a pole in the associated Schr\H{o}dinger operator yields a
special class of weak billiard solutions for nonlinear PDE's.

The soliton fission effect, kink to anti-kink transitions, and multi-peaked
solitons extend to equations that model physical phenomena. The generalized
Kaup equation, the classical Boussinesq system, and the equations governing
second harmonic generation (SHG) are each connected to the cKdV system
through nonsingular transformations. Direct application of these
transformations enables solutions of the cKdV system to be interpreted in
the context of these related equations. Such transformations are given
explicitly in Appendix \ref{app:A}.
Both the Boussinesq System,
\begin{eqnarray} U_t +W_x +UU_x &=& \gamma U_x\ ,\label{eq:bs1a}\\ W_t
+U_{xxx}+(WU)_x &=&
\gamma W_x\ ,\label{eq:bs1b}
\end{eqnarray}
and the generalized Kaup equations
\begin{eqnarray}
  \pi_t &=& \phi_{xx} +{\textstyle\frac{1}{3}} (1 -
  3\sigma)\delta^2\phi_{xxxx} -\epsilon (\phi_x \pi)_x +\alpha
  \pi_x \ , \label{swka}\\
  \phi_t &=& -\alpha \phi_x+{\frac{\epsilon}{2}}\phi_x^2 +\pi \
  ,\label{swkb}
\end{eqnarray}
arise from the theory of shallow water waves (see Whitham(1974)). Here
$\gamma$ and $\alpha$ are small parameters.

In optics, the interaction of a wave envelope at frequency and wavenumber
$(w,k)$ with a second wave at twice the frequency is modeled by the system
of equations
\begin{eqnarray}
  (q_1)_x &=&  -2q_2 q_1^{*} \ ,\label{eq:shginit1a}\\
  (q_2)_\tau &=& q_1^2 \ . \label{eq:shginit1b}
\end{eqnarray}
This process is called second harmonic generation in nonlinear optics
and is used to convert laser light to its second harmonic frequency.

The scattering problem for the energy dependent Sch\H{o}dinger operators was
studied by Jaulent(1972) and Jaulent and Jean(1976).  The completely
integrable variant of the Boussinesq system (\ref{eq:bs1a})-(\ref{eq:bs1b})
was first introduced by Kaup(1972). In Matveev and Yavor(1979)
$\theta$-functions were used to describe quasi-periodic solutions of the
Boussinesq System. They also described a particular type of $N$-soliton
solution using singular classes of $\theta$-functions.  Rational solutions
were studied in Sachs(1998) in connection with a pair of Calogero-Moser
equations coupled through the constraints. Martinez Alonso and Medina
Reus(1992) and Estevez {\it et~al.}(1994) described some of the soliton
solutions using Hirota's $\tau$-functions. Using asymptotics of these
soliton solutions they also demonstrated soliton fission. A connection
between the SHG system and the cKdV system was recently discussed by
Khusnutdinova and Steudel(1998).

\section{Generating Equations for the Coupled KdV and Dym Equations}
\label{sec:one}
We begin by describing the general approach of generating equations and
applying it to the cKdV and cDym equations.  Details for the general case of
$N$-component systems are discussed in Alber {\it et~al.}(1997).

\subsection{Dynamical Generating Equations.}
The hierarchy of the coupled KdV and Dym equations is obtained as the
compatibility condition for the eigenfunction of the linear system of
equations
\begin{eqnarray}
  L \psi &=& 0\ , \label{cmc1}\\
  \psi_t &=& A\psi\ . \label{cmc2}
\end{eqnarray}
The time flow is produced by the linear differential operator
\begin{equation}\label{a}
  A = B\frac{d}{dx} - \frac{B_x}{2}\ ,
\end{equation}
where $B(x,t,E)$ is a specified rational function.  The operator $L$ is
assumed to be of the energy dependent Schr\H{o}dinger type,
\begin{equation}
  L = - \frac{d^2}{dx^2} + V(x,t,E)\ , \label{laz}
\end{equation}
with   a rational potential having the form
\begin{equation} \label{iegenvalue}
{\displaystyle
V(x,t,E) = \frac{\sum_{j = 0}^{N} v_j (x,t) E^j}{\sum_{i = 0}^{M} r_i E^i} },
\end{equation}
where $r_i$ are constants and $v_j (x,t)$ are functions of the variable $x$
and the parameter $t$.  $E$ is a complex spectral parameter.  In particular,
the potential is chosen as
\begin{equation} \label{v}
  V(E)=\kappa E^2 +u(x,t) E +v(x,t)\ ,
\end{equation}
for the cKdV system or
\begin{equation}\label{Vdym}
  V(E)=u(x,t)+E+\frac{v(x,t)}{E} \ ,
\end{equation}
to recover the cDym system. Here $\kappa =\pm 1$. One chooses $\kappa=-1$ to
establish the transformation from the cKdV system to the SHG system, and
$\kappa=1$ to establish the transformation from the cKdV system to the
Boussinesq System.  Notice that the main
difference between the cKdV and cDym systems is the presence of a pole in
the Schr\H{o}dinger operator (\ref{laz}) associated with the cDym equations.
The pole in the potential for the cDym case was shown in Alber {\it
  et~al.}(1994)
to be a necessary feature for systems with weak
billiard solutions.

The compatibility condition of (\ref{cmc1})-(\ref{cmc2}) can be found
by taking the $t$ derivative of (\ref{cmc1}), acting on (\ref{cmc2})
by $L$, and forcing the fact that these two operators commute.  This
leads to the following system of equations
\begin{equation}\label{lazbrac}
 ( L_t + [L,A]) \psi = 0, \;\;\;L \psi = 0\ ,
\end{equation}
where $[L,A]=LA-AL$ is the commutator of $L$ and $A$.
Using the definition of the differential
operator $A$ in (\ref{a}) and $L$ in (\ref{laz}), this Lax equation yields
\begin{equation}\label{gneq}
  \frac{\partial V}{\partial t} = -\frac{1}{2}\frac{\partial^3 B}{\partial
x^3}
  + 2\frac{\partial B}{\partial x}V + B\frac{\partial V}{\partial x}\ ,
\end{equation}
which is a generating equation for the coefficients of the differential
operator $A$. By taking $B$ to be the rational function,
\begin{equation}\label{be}
  B(x,t,E)=\sum_{k=-r}^m b_{m-k}(x,t) \, E^k = E^{-r}\prod_{k=1}^{n}
    (E-\mu_k(x,t))\ ,
\end{equation}
substituting it into the generating equation (\ref{gneq}) and equating like
powers of $E$, a recurrence chain of equations for the coefficients $b_j$ is
obtained. Evaluating these equations one by one, a PDE for the coefficient
$b_{n}$ is obtained, where $n=m+r$.  By considering all possible values of
$n$ and $m$, a hierarchy of systems generated by the Lax equation with a
given potential (generating equation) is  obtained.

Assuming that ${\displaystyle\frac{\partial V}{\partial t}=0}$ in (\ref{gneq})
and integrating gives the stationary generating equation which has the form
\begin{equation} \label{sgneq}
  -B''B +{\textstyle \frac{1}{2}}(B')^2+2B^2 V=C(E)  \ ,
\end{equation}
where the choice of $B(E)$ from (\ref{be})
ensures that $C(E)$ is a rational function with
constant coefficients. These coefficients are the first integrals and
parameters of the coupled system of equations. (For details about the
general method see Alber and Alber(1985).) This equation gives rise to
level sets in the phase space ${\bf C}^n$ corresponding to the Riemann
surface
\begin{equation}
  \label{rims} W^2 = C(E)\ .
\end{equation}
The dynamics of genus $n$ quasi-periodic solutions for $u$ and $v$ with
respect to the $x$ and $t$ coordinates are captured as flows on the level
set produced by a symmetric product of $n$ copies of the Riemann surface
(\ref{rims}) (see Alber {\it et~al.}(1997) for details).

The method of generating equations yields the cKdV and
cDym equations by using $B(x,t,E) =b_0(x,t)E + b_1(x,t)$. In Appendix
\ref{app:A}
solutions
of the generalized Boussinesq and generalized Kaup equations are linked
to these systems. Further, the second harmonic generation equations are
obtained when $B(x,t,E) = b_{2}(x,t)E^{-1}$.

\section{Solutions of the cKdV and cDym Systems from
  Dynamical Systems on Riemann Surfaces}

In this section we obtain finite-dimensional Hamiltonian systems on Riemann
surfaces for the $\mu$ variables defined in (\ref{be}) as
 the roots of the function
$B(x,t,E)$.  These systems capture the
essential dynamics of quasi-periodic and soliton solutions of the cKdV and
cDym systems. The quasi-periodic solutions are often called $n$-gap
solutions in  physics literature and quasi-periodic solutions of genus
$n$ in mathematics literature. In Appendix \ref{app:B} the solutions of these
finite-dimensional Hamiltonian systems are linked to the solutions of the
integrable nonlinear PDE's through trace formulas of the general form
$u = \alpha \sum_{j=1}^n \mu_j + \beta$, where $\alpha$ and $\beta$
are constants. This general
construction also provides links between the solutions of the cKdV and cDym
systems and other nonlinear PDE's.

Choose $B(x,t,E)$ as in (\ref{be}) where $n=r+m$ is the genus of the desired
solution. For the cKdV case, we substitute (\ref{v}) into (\ref{gneq}) and
(\ref{sgneq}) to obtain
\begin{equation}\label{eq}
  u_t E +v_t = -{\textstyle\frac{1}{2}}B'''+2\kappa B' E^2
    +2B'uE+2B'v+Bu'E+Bv'\ ,
\end{equation}
and
\begin{equation}\label{seq}
-B''B + {\textstyle\frac{1}{2}}(B')^2 +2\kappa B^2 E^2 +2B^2uE+2B^2 v=C(E) \ .
\end{equation}
For the cDym system we substitute the potential (\ref{Vdym}) into the same
equations to obtain
\begin{equation}\label{eqdym}
u_tE+v_t=-{\textstyle\frac{1}{2}}B'''E+2B'Eu+2B'E^2+2B'v+Bu'E+Bv' \ ,
\end{equation} and
\begin{equation}\label{seqdym}
  -B''B+{\textstyle\frac{1}{2}}(B')^2+2B^2u+2B^2E+\frac{2B^2v}{E}=C(E) \ .
\end{equation}
By equating like powers of $E$ on the left and right hand side of the
equations (\ref{seq}) and (\ref{seqdym})
we obtain the necessary forms for $C(E)$. Therefore we write
\begin{equation} \label{ce}
  C(E) = {2\kappa}{E^{-2r}}\prod_{i=1}^{2(n+1)} (E-m_i) \ ,
\end{equation}
for cKdV equations and
\begin{equation} \label{ce1}
  C(E) = {2}{E^{-(2r+1)}}\prod_{i=1}^{2(n+1)}(E-m_i)  \ ,
\end{equation}
for the cDym equations, where the real numbers $m_i$ are the roots of the
polynomials
$C(E)$.

\subsection{Finite-Dimensional Hamiltonian Systems.} From the generating
equation above it follows that the $\mu_i$'s, which are the roots of the
function
$B(x,t,E)$, are solutions of finite-dimensional Hamiltonian systems. By
solving
these Hamiltonian systems and using the trace formula, the dynamics of the
roots $\mu_i$ are connected to the functions $u$ and $v$ to obtain solutions
of the PDE's.

Namely, Hamiltonian equations for $\mu_i$'s are obtained by substituting
$E=\mu_i$ into the generating equations (\ref{seq}), (\ref{seqdym}). This
yields the following systems of equations for cKdV:
\begin{equation} \label{mx}
  \mu_i'= \pm \frac{2 \sqrt{\kappa \prod_{j=1}^{2n+2} (\mu_i
      -m_j)}}{\prod_{j\neq i} (\mu_i-\mu_j)} \qquad i=1,\dots,n \ ,
\end{equation}
for the flow in space, and
\begin{equation} \label{mt}
  \dot \mu_i =\mp \frac{2 (\sum_{j\neq i}\mu_j) \sqrt{\kappa
      \prod_{j=1}^{2n+2} (\mu_i-m_j)}}{\prod_{j\neq i} (\mu_i -\mu_j)}
  \qquad i=1,\dots,n \ ,
\end{equation}
for the flow in time. Here $n=m+r$ from the definition of $B$ in (\ref{be})
and $m_i$ are fixed real roots of $C(E)$ in (\ref{ce}).
The plus/minus refers to which branch of the Riemann surface (\ref{rims})
the solution is on.  Systems (\ref{mx})-(\ref{mt})
are commuting Hamiltonian systems with Hamiltonians
\begin{equation}
H=\sum_{j=1}^n \frac{D(\mu_j)(P_j^2-C(\mu_j))}{\prod_{r\neq j}^n
(\mu_j-\mu_r)}
\end{equation}
where $D(\mu_i)=1$ and $D(\mu_i)=-\sum_{j\neq i}\mu_j$ in the
stationary and dynamical cases respectively.  These systems
share the same complete set of first
integrals: $P_j^2 = C(\mu_j), j=1,...,n$ where the polynomial $C(E)$ is
defined
by (\ref{ce}) (for details see Alber {\it et al.}(1997)).

For the cDym equation we see that
\begin{equation} \label{mxdym}
  \mu_i'= \pm \frac{2 \sqrt{ \prod_{j=1}^{2n+2} (\mu_i
      -m_j)}}{\sqrt{\mu_i}\prod_{j\neq i} (\mu_i-\mu_j)} \qquad i=1,\dots,n
  \ ,
\end{equation}
for the spatial flow, and
\begin{equation} \label{mtdym}
  \dot \mu_i =\mp \frac{2 (\sum_{j\neq i}\mu_j) \sqrt{ \prod_{j=1}^{2n+2}
      (\mu_i-m_j)}}{\sqrt{\mu_i}\prod_{j\neq i} (\mu_i -\mu_j)} \qquad
  i=1,\dots,n \ ,
\end{equation}
for the time flow.  Systems (\ref{mxdym})-(\ref{mtdym}) are also
Hamiltonian systems and they share the same complete set of first integrals:
$P_j^2 = C(\mu_j), j=1,...,n$ where polynomial $C(E)$ is defined by
(\ref{ce1}).

Systems (\ref{mx})-(\ref{mt}) and (\ref{mxdym})-(\ref{mtdym}) can be solved
analytically by reducing them to Jacobi inversion problems.  We
will demonstrate the general method in the next section. These equations are
also easily integrated numerically.  Perhaps the best method for
accomplishing numerical integration with the use of symplectic integrators.
Such integrators preserve the Poincare invariants and are stable over a long
period of time, see for example Channell and Scovel(1990).

\subsection{The Trace Formulas.} The connection between the solutions
$u$ and $v$ of
the cKdV equation and the $\mu_i$'s from (\ref{mx}),(\ref{mt})
is derived in Appendix \ref{app:B} and is given by
\begin{eqnarray}
  u&=&2\kappa \sum_{i=1}^{n} \mu_i +2\kappa K_1 \ ,\label{trace1}\\
  v&=&-2\kappa \sum_{i<j\le n} \mu_i \mu_j +\frac{3}{4\kappa} u^2 -K_1 u
  +K_2 \ ,\label{trace2}
\end{eqnarray}
where
\begin{eqnarray}
  K_1 &=& -\frac{1}{2} \sum_{i=1}^{2n+2} m_i \ ,\\
  K_2 &=& \kappa\sum_{i<j\le 2n+2} m_i m_j -\kappa K_1^2 \ .
\end{eqnarray}
Observe that the small parameter $K_1$ from (\ref{cKdVinit})
is zero if $\sum_{i=1}^{2n+2} m_i=0$.
The trace formulas for the cDym system are
\begin{eqnarray}
  u&=&2 \sum_{i=1}^{n} \mu_i +2K_1 \ ,\\
  v&=&-\frac{1}{4}u''-2 \sum_{1\le i<j\le n} \mu_i \mu_j
  +\frac{3}{4}u^2 -K_1
  u+K_2 \ .
\end{eqnarray}

\section{Classification of Limits of Periodic Solutions}
For the next three sections we seek various periodic (genus-one)
solutions of the Boussinesq System.
Therefore unless otherwise stated we assume $r=0$ from (\ref{be})
and $\kappa=1$ from (\ref{v}).  We first obtain the periodic
traveling-wave solutions and then show that they are equivalent
to solutions obtained  in
terms of a $\mu$ variable in the case $n=1$. This provides a natural
introduction to the algebraic geometric method.  A one-soliton
solution is then obtained by deforming the Riemann surface of the genus-one
periodic solution.  This method of first finding
periodic/quasi-periodic solutions and then
deforming the level set (Riemann surface)
in the phase space
to obtain soliton solutions will be utilized throughout this paper.
(For details about general approach see amongst others Ablowitz and
Ma(1981)
and Alber and Alber(1985).)

\subsection{Periodic Traveling-Wave Solutions} Let $U=U(\zeta)$ and
$W=W(\zeta)$ where $\zeta=x-ct$
so that the Boussinesq System becomes
\begin{eqnarray}
  -cU' +W' +UU' &=& \gamma U' \   ,\label{bsprime1}\\
  -cW' +U'''+(WU)' &=& \gamma W'\ , \label{bsprime2}
\end{eqnarray}
where $W'$ and $U'$ denote differentiation with respect to $\zeta$. Then
(\ref{bsprime1}) gives
\begin{equation}\label{W}
  W=\eta U-{\textstyle \frac{1}{2}}U^2+\tau_0\ ,
\end{equation}
where $\eta=c+\gamma$ and $\tau_0$ is a constant of integration. Plugging
this into (\ref{bsprime2}) and integrating twice gives
\begin{equation}
  \frac{\tau_0-\eta^2}{2}U^2+\frac{\eta}{2}U^3+\frac{1}{2}(U')^2-\frac{1}{8}
  U^4=\tau_1 U +\tau_2\ ,
\end{equation}
where $\tau_1,\tau_2$ are constants of integration. Writing this as an
integral equation and taking a square root we obtain
\begin{equation}\label{dzeta}
  d\zeta = \pm \frac{dU}{\sqrt{C_4(U)}}
\end{equation}
where
$$C_4(U) = \prod_{l=1}^4 (U - m_j)= U^4-4\eta
U^3-4(\tau_0-\eta^2)U^2-8\tau_1U-8\tau_2.\label{eq:c4u}
$$
Notice that the right hand side of this differential equation is multi valued
since it involves a square root. This is uniquely defined on a
Riemann surface of genus $1$ parametrized by a pair $(W,E)$ where
\begin{equation} \label{r62}
  W^2 = C_4(E) = \prod_{l=1}^4 (E - m_j).
\end{equation}
One indicates one of two sheets of the Riemann surface by choosing a
particular sign in front of the
square root: $W = \pm \sqrt{C_4(E)}$.
Therefore $U$ in (\ref{dzeta}) is considered on a particular sheet of the
Riemann surface (\ref{r62})  and so meaningful integration can take place.

The equation (\ref{dzeta}) can  also be obtained from the $\mu$-equations
(\ref{mx}) and (\ref{mt})
for $n=1$. Notice that the trace formula shows that $U$ and $\mu$ are
linearly related so that one can substitute $\mu$ instead of $U$ in
(\ref{dzeta}). Using $\mu$ in this
case will make it consistent with the formulas in the case when $n=2$ to be
described in the next
section.

\begin{figure}
\begin{center}
\includegraphics[height=3in]{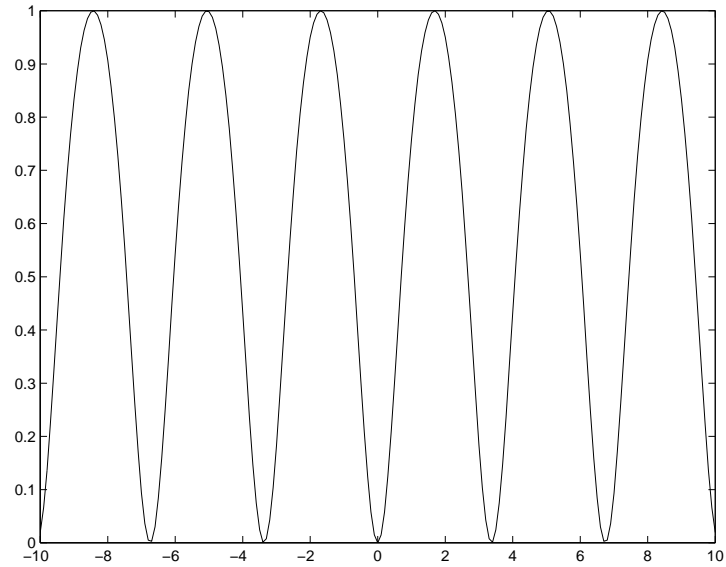} \end{center}
\caption{\label{figper} Periodic $\mu$ solving (\ref{dmu}) at $t=0$ in
the genus-1 case. This corresponds
to distinct roots $m_i$ in (\ref{ce}). This plot uses
$m_1=0,\;m_2=3,\;m_3=2,\;m_4=1$.  From this function $\mu$,
solutions $u,v$ of the cKdV equation may be obtained by
(\ref{trace1}),(\ref{trace2}).  Then $U,W$ from the Boussinesq equation
may be found using (\ref{coupledKdV1}) and (\ref{coupledKdV2}).}
\end{figure}

After integrating (\ref{dzeta}), the following Jacobi
inversion problem is obtained
\begin{equation}\label{dmu}
 \theta=  x-ct+\theta_0 = \frac{1}{2}\int_{\mu_0}^{\mu}
      \frac{d\mu}{\sqrt{\prod_{i=1}^4 (\mu-m_i)}}\ ,
\end{equation}
As stated before, this is a typical
Jacobi inversion problem (Mumford (1983)). This integral
is inverted using Jacobi's elliptic functions,
\begin{equation}\label{firstkind}
 \frac{\sqrt{(m_1-m_3)(m_2-m_4)}}{2} \int_{\mu_0}^{\mu} \frac{
    d\mu}{\sqrt{\prod_{i=1}^4(\mu-m_i)}} = \int_{z_0}^z
  \frac{dz}{\sqrt{(1-z^2)(1-k^2z^2)}} \ ,
\end{equation}
where
\begin{equation}
  z^2=\frac{(m_2-m_4)(\mu-m_1)}{(m_1-m_4)(\mu-m_2)} \; ,\;
 k^2 = \frac{(m_1-m_4)(m_2-m_3)}{(m_2-m_4)(m_1-m_3)} \ .
\end{equation}
Notice that (\ref{firstkind}) is an elliptic integral of the first kind (see
Mumford (1983)).  Therefore, $\mu$ is obtained explicitly and
\begin{equation}
  \mu(x,t)=\frac{m_2(m_4-m_1){\rm sn}^2(k,\omega)+m_1(m_2-m_4)}
  {m_2-m_4+(m_4-m_1){\rm sn}^2(k,\omega)} \ ,
\end{equation}
where ${\rm sn}(k,\omega)$ is the Jacobi ${\rm sine}$ function and
$\omega=(x-ct+\theta_0)/\sqrt{(m_1-m_3)(m_2-m_4)}$. This function is plotted
in Figure~\ref{figper}.

Periodic solutions correspond to the case when all $m_i$ are distinct. In
this case, each fixed point in the phase space repels the
trajectories so that for any initial value, $\mu$ oscillates periodically
between the two nearest  points.  Notice that $\mu(0)$ must be
chosen so that the right hand side of (\ref{dmu}) is real.  Then the fixed
points repel  and $\mu$ remains real valued.  The
solution $\mu$ leads
to solutions for $U$ and $W$ through the trace formulas. The shape of $U$ is
essentially the same as $\mu$. The shape of $W$ is aperiodic and is
discussed further in the next section. Equation (\ref{dmu}) can be
interpreted as  defining an angle variable $\theta$ where $c$ is then the
action variable. From (\ref{dmu}) it follows that in terms of these variables
the initial Hamiltonian flow linearizes. It also can be viewed as an Abel
-Jacobi map
from a hyperelliptic curve
\begin{equation}
  W^2 = C(E) = 2(\mu - m_1)(\mu - m_2)(\mu - m_3)(\mu - m_4)\ ,
\end{equation}
or in other words, a Riemann surface of genus one, onto the Jacobi variety:
$J= [{\bf C}|w Z]$ where ${\bf C}$ is a complex plane and $w$ is the
period lattice of the holomorphic differential from (\ref{dmu}).

\subsection{One-Soliton Solution of Kaup Type}
To examine soliton solutions, one deforms the Riemann surface $W^2=C_4(E)$.
(For details about soliton deformations see amongst others Ablowitz and
Ma(1981)
and Alber and Alber(1985).) Here we consider the limit $m_1
\rightarrow m_2 \rightarrow a$, i.e. where the 2 roots $m_1$ and $m_2$
coalesce into one point.  As this limit is approached,  a soliton solution is
obtained as the period of the periodic solution
increases to infinity. In this case  $a$ is called a
double point. This double point is an attractor in the phase space. Without
loss of generality, we assume $a <m_3 <m_4$. This assumption leads to the
realization that the only pertinent solution is obtained when
$ a < \mu(0) <m_3$.

\begin{figure}
\begin{center}
\includegraphics[height=3in]{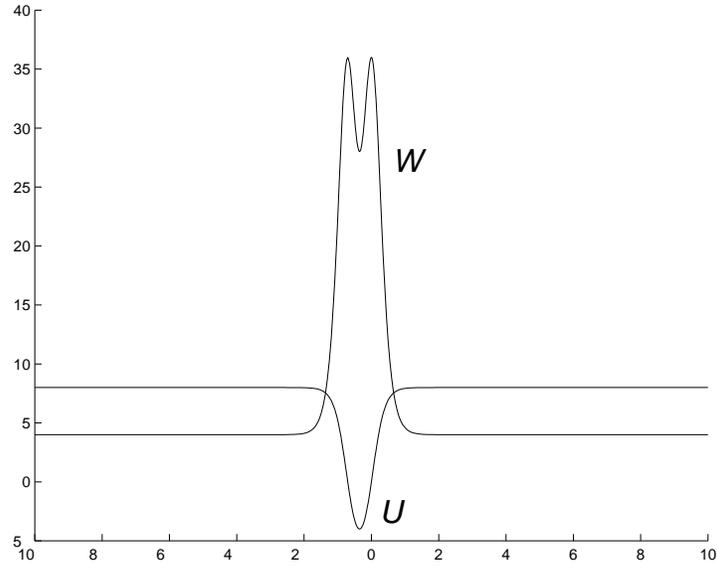}
\end{center}
\caption{\label{fig:2hump}Genus-one solutions $U$ and $W$ to the
Boussinesq equation obtained from (\ref{U1}) using $a=-2, m_3=1,m_4=3$.}
\end{figure}

 On the principle branch
of the square root,
\begin{eqnarray}
  x-ct+\theta_0 &=& \frac{1}{2}\int_{\mu_0}^{\mu} \frac{
    d\mu}{(\mu-a)\sqrt{(\mu-m_3)(\mu-m_4)}} \ ,\\
  &=&  -\left[(a-m_4)(a-m_3)\right]^{-1/2}
  {\rm arctanh}\left(\sqrt{\frac{(a-m_4)(\mu -m_3)}{(a-m_3)(\mu
    -m_4)}}\right)
  \ ,
\end{eqnarray}
so that
\begin{equation}
  \mu=\frac{m_3(m_4-a)+m_4(a-m_3) {\rm tanh}^2(\omega)}{(m_4-a)+
    (a-m_3){\rm tanh}^2(\omega)} \ ,
\end{equation}
where $\omega=-(x-ct+\theta_0 ) \sqrt{(a-m_4)(a-m_3)}$ and $a,m_3,m_4$ are
functions of $c$. Using the trace formulas (\ref{trace1})-(\ref{trace2})
and the transformations
(\ref{coupledKdV1})-(\ref{coupledKdV2}), the exact formulas for $U$ and
$W$ from the Boussinesq equation are obtained.  For example,
\begin{equation}\label{U1}
U=\frac{-4a^2+(m_3-m_4)^2+2a(m_3+m_4)+(m_4^2-m_3^2){\rm cosh}
(2\omega)} {2a-(m_3+m_4)+(m_3-m_4){\rm cosh}(2\omega)} \ .
\end{equation}
 Notice that $\mu$ has a shape similar to a KdV
solitary wave. It has a peak at $m_3$ and approaches $m_1$ for large $|x|$.
 From this we conclude that $U$ is also shaped like a KdV soliton with a peak
of height
$4a+2m_4-2m_3$ and for large $|x|$ approaches $2m_3+2m_4$. One might be
surprised that the soliton does not approach 0 as $x\rightarrow \pm \infty$.
But remember that this is the modified cKdV equation with $K_1\neq 0$.
Howver we may choose the parameters so that the solution does approach
zero.   The solution is plotted in Figure~\ref{fig:2hump}. This
solution was first found by Kaup(1972) using the inverse scattering
transform in case when
$m_3=-m_4$ and
$a=0$, hence $K_1=0$.
One advantage of the approach used here is that $W$ is easily found
using the trace formulas, and the relationship between $U$ and $W$ is seen
explicitly.

\subsection{Solutions with Two Peaks.} Observe next that since $n=1$, $W$ is
only a quadratic in $U$. The polynomial
\begin{equation}
  W=-{\textstyle\frac{1}{2}}U^2+(2a+m_3+m_4) U -4a m_3 -4 a m_4 +(m_3
  -m_4)^2 \ ,
\end{equation}
obtained from the trace formula is a parabola in $U$ with vertex at $U=2a
+m_3+m_4$. Since $U$ has the shape of a solitary wave, $W$ has either one or
two peaks depending on the parameters defining the vertex of $U$.
(See Figure~\ref{fig:2hump}). If $U$ is concave down,
then $W$ is a double peaked
soliton if $m_3+m_4<2a$ and $3m_3<2a+m_4$.  In the unperturbed case
($\gamma=0$), these conditions reduce to $m_3<0<a$.  Otherwise $W$ has a
single peak.  The concave up case is similar.

\subsection{One Kink Solution}
The one-kink solutions were initially found by Alonso and Rues(1992) as the
simplest solutions obtained using the bilinear formalism of the Kyoto
school.  Our method introduces an alternate description of soliton fusion
and fission for $n=2$ and simplifies the computations.

\begin{figure}
\begin{center}
\includegraphics[height=3in]{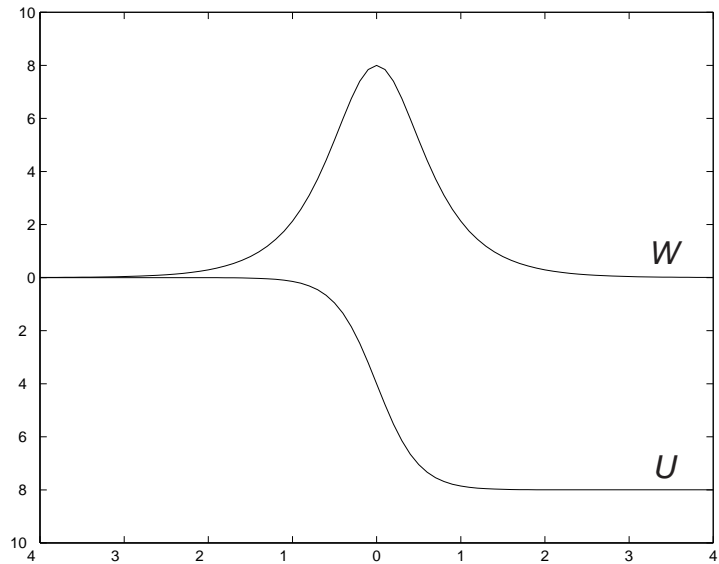} \end{center}
\caption{$U$ and $W$ shown here are
genus-one solutions to
the Boussinesq equation for two double roots. $t=0$,
$\theta_0=0$, $c=-4$.  This graph highlights the simple relationship
between $U$ and $W$ which will be important in understanding two
soliton interaction.}
\label{fig:cbcase}
\end{figure}

To obtain the one-kink solutions, we deform the Riemann surface
further by taking the limit $m_3 \rightarrow m_4$, so that $ m_1=m_2=a_1$
and $m_3=m_4=a_2$.  Note that if $\kappa=-1$ this limit is not permitted
for real valued
$\mu$ in (\ref{mx}).  However, for $\kappa=1$
the system is readily solved for real $\mu$
producing a special case of  equation (\ref{dzeta}) where
$\tau_1=0=\tau_2=\gamma$. After inverting, the angle variable is
\begin{eqnarray}
  \theta=x-ct+\theta_0 &=& \pm \int \frac{ 2dU}{U(U-2c)} = \mp \frac{ 2}{c}
{\rm arctanh}\left( \frac{U-c}{c} \right) \ ,
\end{eqnarray}
and we find using the trace formulas and the transformations obtained
in
Appendix \ref{app:A} that
\begin{eqnarray}
  U &=& \pm c \; {\rm tanh}(-\frac{c}{2} (x-ct+\theta_0)) +c_2 \ ,\\
  W &=& \pm\frac{c^2}{2}{\rm sech}^2(-\frac{c}{2}(x-ct+\theta_0)) \ .
\end{eqnarray}
These functions are plotted in
Figure~\ref{fig:cbcase}.  Here $U$ is always a kink or anti-kink while $W$
is always similar to a typical KdV soliton. Observe that the speed of the
soliton, $c^2/2$ is exactly the amplitude of $W$ and is proportional to the
square of the amplitude of the $U$ soliton. This connection between wave
speed and amplitude is reminiscent of that found in the KdV equation.

\section{Classification of Limits of Genus-Two Solutions}
In this section genus-two quasi-periodic solutions are constructed.  By
deforming the Riemann surface (spectral polynomial), two-soliton solutions
are obtained and several types of soliton-soliton interactions are
described.

\subsection{Quasi-periodic Genus-Two Solutions}
In the genus-two case $U$ is given by the trace formulas to be
the sum of two periodic functions, $\mu_1$ and $\mu_2$, and a constant.  In
general the functions, $\mu_1$ and $\mu_2$, have noncommensurate periods, so
that
the solution $U$ is generically quasi-periodic and systems (\ref{mx}) and
(\ref{mt}) are defined on the symmetric product of two copies of the Riemann
surface (hyperelliptic curve) of genus two given by
\begin{equation} \label{r6}
  W^2 = C_6(E)\ ,
\end{equation}
where
\begin{equation}
  C_6(E) = \kappa \prod_{l=1}^6 (E - m_j)\ .
\end{equation}
After reordering the equations, summing them up, and integrating, the
following Jacobi inversion problem is obtained:
\begin{eqnarray}
  \theta_1 &=& \int_{\mu_1^0}^{\mu_1} \frac{d \mu_1}{\sqrt{C_6(\mu_1)}} +
  \int_{\mu_2^0}^{\mu_2} \frac{d \mu_2}{\sqrt{C_6 (\mu_2)}} = 2x - 2 a_1 t +
  \theta_1^0\label{an0a}\ ,\\
  \theta_2 &=&\int_{\mu_1^0}^{\mu_1} \frac{\mu_1 d \mu_1}{\sqrt{C_6
      (\mu_1)}} + \int_{\mu_2^0}^{\mu_2} \frac{\mu_2 d
    \mu_2}{\sqrt{C_6(\mu_2)}} = 2x -2 a_2 t + \theta_2^0 \ . \label{an0b}
\end{eqnarray}
Inverting the Abel-Jacobi map defined by (\ref{an0a})-(\ref{an0b}) results
in expressions for $\mu_1$ and $\mu_2$ in terms of Riemann
$\theta$-functions. (For details about the Abel-Jacobi map see
Mumford(1983), Matveev and Yavor(1979),
and  Ercolani and McKean (1990).)

Having derived the genus-two quasi-periodic solutions, several limiting cases
will now be explored to introduce solitons. Below each distinct case is
considered.

\subsection{One-Soliton Solution on a Quasi-periodic Background}
\begin{figure}
\begin{center}
\includegraphics[height=3in]{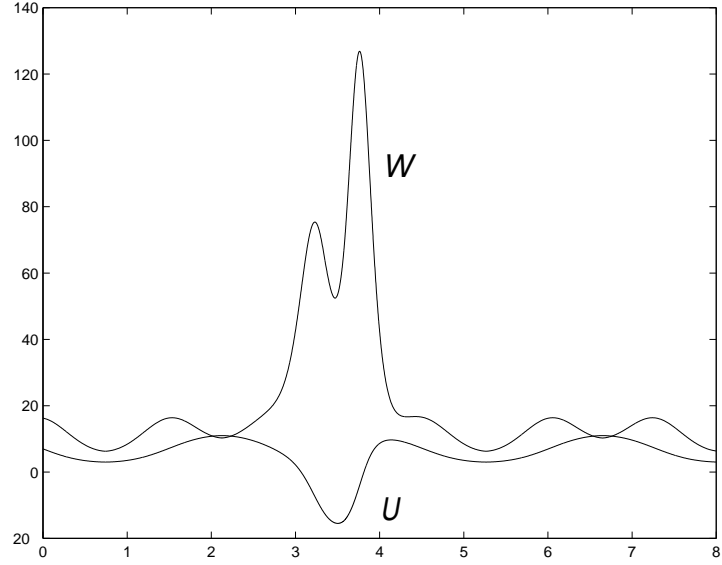} \end{center}
\caption{\label{fig:quasiback}Genus-two solutions $U$ and $W$
to the Boussinesq equation for one double root derived from
(\ref{2an1a}), (\ref{2an1b}) and the trace formulas.
This corresponds to a Kaup type soliton on a quasi-periodic background.
Here $t=0$, $a=-5$,$m_3$=0,$m_4=1$,$m_5=3$,$m_6=3.5$,
$\mu_1(0)=-4.99$, $\mu_2(0)=2$.}
\end{figure}

A one-soliton solution on a quasi-periodic background is obtained in the
limit $m_1\rightarrow m_2\rightarrow a$.  Just as in the one-soliton case, a
soliton is created as the Riemann surface is manipulated by pinching two
elements of the spectrum. However, in this case there are two $\mu$ variables
and the orbit for only one $\mu$-variable is changed. The other $\mu$ variable
remains periodic.  The result is a  solution with a Kaup
type soliton on a quasi-periodic background, and it is plotted in
Figure~\ref{fig:quasiback}.
The problem of inversion may be written in the following way,
\begin{eqnarray}
  \theta_1 &=& \int_{\mu_1^0}^{\mu_1} \frac{d \mu_1}{(\mu_1 - a)
    \sqrt{P_4(\mu_1)}} + \int_{\mu_2^0}^{\mu_2} \frac{d \mu_2}{(\mu_2 - a)
    \sqrt{P_4(\mu_2)}} = 2 t + \theta_1^0
  \ , \label{2an1a}\\
  \theta_2 &=& \int_{\mu_1^0}^{\mu_1} \frac{d \mu_1}{\sqrt{P_4(\mu_1)}} +
  \int_{\mu_2^0}^{\mu_2} \frac{d \mu_2} {\sqrt{P_4(\mu_2)}} = 2x - 2a t +
  \theta_2^0 \ , \label{2an1b}
\end{eqnarray} where
\begin{equation}
  P_4(E) = (E - m_3)(E - m_4)(E - m_5)(E - m_6)\ .
\end{equation}
(For details about inverting problems of this type see
Alber and
Fedorov(1999).)

\subsection{Two-Soliton Solutions of Kaup Type}

The two-soliton solution is obtained by piecewise
pinching together two pairs of elements
of the spectrum so that $m_1\rightarrow m_2 \rightarrow a_1$ and $m_3
\rightarrow m_4\rightarrow a_2$. Here there are two double points $a_1$ and
$a_2$, as well as two remaining hyperelliptic points at $m_5$ and $m_6$.
Choosing
the initial data on the positive branch of the Riemann surface for both
$\mu$ variables leads to the following
problem of inversion
\begin{eqnarray}
  \label{an1}\nonumber
  \theta_1 &=& \int_{\mu_1^0}^{\mu_1} \frac{d \mu_1}{(\mu_1 - a_2)
    \sqrt{(\mu_1 - m_5)(\mu_1 - m_6)}}\nonumber \\ &&+
  \int_{\mu_2^0}^{\mu_2} \frac{d \mu_2}{(\mu_2 - a_2) \sqrt{(\mu_2 -
      m_5)(\mu_2 - m_6)}} = 2x - 2 a_1 t + \theta_1^0 \ ,\\
  \label{an1part2}
  \theta_2 &=&\int_{\mu_1^0}^{\mu_1} \frac{d \mu_1}{(\mu_1 -a_1)
    \sqrt{(\mu_1 - m_5)(\mu_1 - m_6)}} \nonumber \\ &&+
  \int_{\mu_2^0}^{\mu_2} \frac{d \mu_2}{(\mu_2 - a_1) \sqrt{(\mu_2 -
      m_5)(\mu_2 - m_6)}} = 2x -2 a_2 t + \theta_2^0 \ .
\end{eqnarray}
This angle representation is similar to the one found in the case of the
defocusing NLS equation. (See Alber and Marsden(1994).)

Notice that the $\theta_i$'s are essentially the sum of two Kaup type solitons
as $t \rightarrow \pm \infty$.   The
integrals in system (\ref{an1}),(\ref{an1part2})
may be evaluated to obtain the following
nonlinear algebraic system of equations for the $\mu_i$:
\begin{eqnarray}
  (s_{1625}+s_{1526})(s_{2625}+s_{2526})&=&
  A_1(s_{1625}-s_{1526})(s_{2625}-s_{2526})\ ,\\
  (s_{1615}+s_{1516})(s_{2615}+s_{2516}) &=&
  A_2(s_{1615}-s_{1516})(s_{2615}-s_{1516})\ ,
\end{eqnarray}
where
$s_{ijkl} = \sqrt{(\mu_i-m_j)(a_k-m_l)}$ and
\begin{eqnarray}
  A_1&=&\exp[-\sqrt{(a_2-m_5)(a_2-m_6)}(2x-2a_1t+\theta_1^0)] \ ,\\
  A_2&=&\exp[-\sqrt{(a_1-m_5)(a_1-m_6)}(2x-2a_2t+\theta_2^0)] \ .
\end{eqnarray}

\subsection{Phase Shift Formulas} When two solitons interact, they normally
re-emerge with their initial profile and velocity. However, they
have shifted ahead or behind where they would have been had there been no
interaction at all.  The amount a soliton shifts is called its
phase shift, and the
integral equations (\ref{an1}),(\ref{an1part2}) can be used to compute it.
Assume that $a_1>a_2$ and define
\begin{equation}
  M(x)=\sqrt{\frac{x-m_5}{x-m_6}} \ .
\end{equation}
The phase shifts for the systems (\ref{an1})-(\ref{an1part2})
are calculated by using the
following procedure. First, consider the reference frame where $\theta_1$
is a constant, that is $2x-2a_1t+\theta_1^0=\alpha_1$ for some constant
$\alpha_1$. Then observe that $\theta_2$ can be written only as a function
of $t$ and $\alpha_1$. Namely, $\theta_2 = \alpha_1 + 2(a_1-a_2)t +
\theta_2^0 - \theta_1^0$. Notice that the integrals on the left hand side of
(\ref{an1}) are exactly the expression for single Kaup type solitons.
Integrals of this form may be integrated as
\begin{equation}
  \int_{\mu_2^0}^{\mu_2} \frac{d \mu_2}{2(\mu_2 - a_1) \sqrt{(\mu_2 -
      m_5)(\mu_2 - m_6)}} = \frac{1}{2\sqrt{(a_1-m_5)(a_1-m_6)}} \log \left|
    \frac{\psi-M(a_1)}{\psi+M(a_1)} \right|\ ,
\end{equation}
where for clarity we have defined $\psi^2=M(\mu_2)^2$.  Then we see that in
this frame when $t\rightarrow\infty$ the right side of (\ref{an1part2})
goes to infinity so that the left hand side must also grow without bound. On
evaluation of the integrals we see that this can only happen when $\psi
\rightarrow -M(a_1)$. Similarly when $t\rightarrow -\infty$, $\psi$ must
approach $M(a_1)$. Substituting these values for $\mu_2$ in (\ref{an1part2})
gives asymptotics for $\theta_1$ as $t\rightarrow\infty$ and as
$t\rightarrow -\infty$, respectively. The phase shift for one of the
solitons will be the difference between the behavior of $\theta_1$ at minus
infinity and its behavior at plus infinity in the frame given by
$\theta_1=\alpha_1$. In this way we obtain the shift between phases before
and after interaction for $\theta_1$ to be
\begin{equation}\label{Delta1}
  \Delta_1=\frac{1}{\sqrt{(a_2-m_5)(a_2-m_6)}} \log \left
    |\frac{M(a_2)+M(a_1)}{M(a_2)-M(a_1)} \right| \ .
\end{equation}
Similarly the phase shift for $\mu_2$ along $\theta_2=\alpha_2$ is computed
to be
\begin{equation}\label{Delta2}
  \Delta_2=\frac{1}{\sqrt{(a_1-m_5)(a_1-m_6)}} \log \left |
    \frac{M(a_1)+M(a_2)}{M(a_1)-M(a_2)} \right| \ .
\end{equation}
These formulas were initially obtained by Matveev and Yavor(1979) by
studying asymptotics of singular $\theta$-functions.

\subsection{Kink-Anitkink Interaction Solutions}
The interaction of kink and antikink solutions will now be considered. The
two-kink solutions are constructed by taking the limit of the spectral
parameters so that $m_1\rightarrow m_2 \rightarrow a_1$, $m_3 \rightarrow
m_4 \rightarrow a_2$, and $m_5\rightarrow m_6 \rightarrow a_3$. We examine
both the case when the $\mu_i$'s are both initially on the positive branch of
the Riemann surface and the case when only one of the $\mu_i$'s is initially
on the positive branch and the other is on the negative branch.  In contrast
to the KdV equation, this
difference in the initial conditions produces qualitatively different
solutions. This difference arises because
the KdV and cKdV equations contain at least one
hyperelliptic branch point, while all of the fixed points are double points
for the two-kink solutions.

\subsection{Initial values of $\mu_1$ and $\mu_2$ are chosen on the
positive branches of the Riemann
surface.} In this situation, the angle variables are computed to be
\begin{eqnarray} \label{an22} \theta_1 &=& +\int_{\mu_1^0}^{\mu_1} \frac{d
    \mu_1} {(\mu_1 - a_2) (\mu_1-a_3)} \nonumber\\ & &+
  \int_{\mu_2^0}^{\mu_2} \frac{d \mu_2}{(\mu_2 - a_2) (\mu_2 - a_3)} = 2x -
  2a_1 t + \theta_1^0 \ ,\\
  \theta_2 &=&+\int_{\mu_1^0}^{\mu_1} \frac{d \mu_1}{(\mu_1 - a_1) (\mu_1 -
    a_3)} \nonumber\\
  & &+ \int_{\mu_2^0}^{\mu_2} \frac{d \mu_2}{(\mu_2 - a_1)(\mu_2 - a_3)} =
  2x - 2 a_2 t + \theta_2^0 \ .
\end{eqnarray}
These integrals are tractable, and the
symmetric polynomials $\mu_1\mu_2$ and $\mu_1+\mu_2$ which appear in the trace
formulas, can be calculated explicitly from the resulting system
\begin{eqnarray}
  (1-A_1)\mu_1\mu_2+(a_3A_1-a_2)(\mu_1+\mu_2) &=& A_1 a_3^2-a_2^2 \ ,
  \label{al1a} \\
  (1-A_2)\mu_1\mu_2+(a_3A_2-a_1)(\mu_1 +\mu_2) &=& A_2 a_3^2-a_1^2 \ ,
  \label{al1b}
\end{eqnarray} with
\begin{eqnarray}\label{mu1plusmu2}
  A_1=\frac{(\mu_1^0 -a_2)(\mu_2^0 -a_2)}{(\mu_1^0
    -a_3)(\mu_2^0 -a_3)} \, e^{2(a_2-a_3)(x-a_1t+\theta_1^0)} \ ,\\
  A_2=\frac{(\mu_1^0 -a_1)(\mu_2^0 -a_1)}{(\mu_1^0 -a_3)(\mu_2^0
    -a_3)} \, e^{2(a_1-a_3)(x-a_2t+\theta_1^0)} \ .
\end{eqnarray}
These calculations are carried out for the unperturbed Boussinesq
equations ($K_1=0$) in the
next section and soliton fusion and fission are discussed.

\subsection{Soliton Fusion and Fission.} Given this particular deformation of
the Riemann surface, solitons can undergo fission or fusion. This interesting
phenomenon occurs when two separate solitons enter an interaction but only
one single soliton emerges from this interaction.
Since all of the equations under study are invariant under the
space-time inversion
$(x\rightarrow -x, t\rightarrow -t)$,
the reverse process of soliton fission may also occur
where a single soliton breaks into two distinct solitons at some critical
time.  Observe that soliton fission can be interpreted as an infinite
phase shift.  This is because the solitons change their speed since when they
fuse together.  Therefore they will be infinitely far from where they would
have been had there been no interaction.
In this sense formulas (\ref{Delta1})-(\ref{Delta2}) are still
correct since in the limit $m_5=m_6=a_3$, $M(x)=1$ and so the formulas become
singular.

For this discussion we assume that $\gamma=-2\sum_{i=1}^{2n+2} m_i=0$
although the general case is similar. Hence the system
(\ref{al1a})-(\ref{al1b}) can be solved for $\mu_1+\mu_2$ and is found to
be
\begin{equation}\label{burgerlike}
U=-4(\mu_1+\mu_2)=-4 \frac{a_1f_1+a_2f_2+a_3f_3}{f_1+f_2+f_3}\ ,
\end{equation}
where
\begin{eqnarray}
f_1&=&\exp(2a_1x+2a_1^2t+\tilde{\theta_1}) \ ,\\
f_2&=&\exp(2a_2x+2a_2^2t+\tilde{\theta_2}) \ ,\\
f_3&=&\exp(2a_3x+2a_3^2t+\tilde{\theta_3}) \ ,
\end{eqnarray}
and
\begin{eqnarray}
\tilde{\theta_1}&=&
a_1\theta_1^0+\log{[(a_2-a_3)(\mu_1^0-a_1)(\mu_2^0-a_1)]} \ ,\\
\tilde{\theta_2}&=&
a_2\theta_1^0+\log{[(a_3-a_1)(\mu_1^0-a_2)(\mu_2^0-a_2)]} \ ,\\
\tilde{\theta_3}&=&
a_3\theta_2^0+\log{[(a_1-a_2)(\mu_1^0-a_3)(\mu_2^0-a_3)]} \ .
\end{eqnarray}

The expression (\ref{burgerlike}) coincides with a solution to Burger's
equation which describes a confluence of shocks (see
Whitham(1974)) and has
the same form as a solution obtained by using the Hirota method. This
expression is now analyzed to see why it represents
soliton fusion.  Suppose $a_1<0<a_2<a_3$ and
consider $t=0$. We claim that at this instant (\ref{burgerlike}) is a
two-tiered kink (sum of two kinks), see Figure ~\ref{fig3cd}.  To see why this
is the case, consider $2x<(\tilde{\theta_2}-\tilde{\theta_1})/(a_1-a_2)$.
For these values of $x$, $f_1$ is the largest of the terms $f_1$, $f_2$, and
$f_3$. In fact as $x\rightarrow -\infty$, the terms $f_2$,$f_3$ are
negligible in comparison to $f_1$. So for these values of $x$, $\mu_1+\mu_2$
is nearly constant and is approximately $a_1$. Similarly for
$(\tilde{\theta_2}-\tilde{\theta_1})/(a_1-a_2)<2x<
(\tilde{\theta_3}-\tilde{\theta_2})/(a_2-a_3)$, $f_2$ is the dominant term
and $\mu_1+\mu_2\approx a_2$. For the remaining values of $x$, $f_3$
dominates so $\mu_1+\mu_2\approx a_3$.

We now apply this analysis for arbitrary $t$. In this manner we get that $U$
is essentially constant in three regions, $D_1$, $D_2$ and $D_3$, of the
$(x,t)$ plane as seen in Figure ~\ref{fusion}. When $t$ is sufficiently small,
these regions are bounded by the points where the functions $f_1=f_2$ and
where $f_2=f_3$. At
\begin{equation}
  t^*=\frac{(a_2-a_3)(\tilde{\theta_2}-\tilde{\theta_1})-(a_1-a_2)
    (\tilde{\theta_3}-\tilde{\theta_2})}{2(a_1-a_2)(a_2-a_3)(a_1-a_3)}\ ,
\end{equation} the
functions $f_1=f_2=f_3$
for some $x^*$. From this instant on, $f_2$
is never the largest term and so for $t>t^*$, the plane is divided
into two regions bounded by the points where $f_1=f_3$ as seen in
Figure~\ref{fusion}. For more details on this analysis see
Whitham's(1974) chapter on
Burger's equation. As explained above, the regions $D_i$, which contain all
the information regarding fission of the two-kink solution (times, speeds
etc.), are derived from the parameterization of the lines $f_1=f_2$,
$f_2=f_3$,
and $f_1=f_3$.  These lines are computed explicitly giving that $f_1=f_2$
along
the line where $2x=2a_3t +(\tilde{\theta_2}-\tilde{\theta_1})/(a_1-a_2)$,
$f_2=f_3$ along the line where $2x=2a_1t
+(\tilde{\theta_3}-\tilde{\theta_2})/(a_2-a_3)$ and $f_1=f_3$ along the
lines where $2x=2a_2t+(\tilde{\theta_3}-\tilde{\theta_1})/(a_1-a_3)$. From
this it is possible to predict the times at which two solitons
experience fission or
fusion based solely on the initial phase and
the values of the three spectrum points
$a_i$. The speeds of the solitons are given by the slopes of the lines, and
in this way a complete explanation of fission is given.  Notice that
the arguments are quite general and that this approach can be applied
directly to the entire class of $N$-component systems.

\subsection{Initial values of $\mu_1$ and $\mu_2$ are chosen on different
branches of the Riemann surface.} This case gives a similar
problem of inversion as the fusion case, namely
\begin{eqnarray}
  \theta_1 &=& +\int_{\mu_1^0}^{\mu_1} \frac{d \mu_1} {(\mu_1 - a_2)
    (\mu_1-a_3)} \nonumber\\ & &- \int_{\mu_2^0}^{\mu_2} \frac{d
    \mu_2}{(\mu_2 - a_2) (\mu_2 - a_3)} = 2x -
  2a_1 t + \theta_1^0 \ ,\label{an2a}\\
  \theta_2 &=&+\int_{\mu_1^0}^{\mu_1} \frac{d \mu_1}{(\mu_1 - a_1) (\mu_1 -
    a_3)} \nonumber\\
  & &- \int_{\mu_2^0}^{\mu_2} \frac{d \mu_2}{(\mu_2 - a_1)(\mu_2 - a_3)} =
  2x - 2 a_2 t + \theta_2^0 \ .\label{an2b}
\end{eqnarray} These integrals may be evaluated in terms of the $\log$
function and give rise to the following algebraic equations for the $\mu_i$'s
\begin{eqnarray}\label{systemofmui}
  \frac{(\mu_1-a_1)(\mu_2-a_3)}{(\mu_1-a_3)(\mu_2-a_1)} &=&
  \frac{(\mu_1^0-a_1)(\mu_2^0-a_3)}{(\mu_1^0-a_3)(\mu_2^0-a_1)}
  \exp[(a_1-a_3)(2x-2a_2t+\theta_2^0)]  \ ,\\
  \frac{(\mu_1-a_2)(\mu_2-a_3)}{(\mu_1-a_3)(\mu_2-a_2)}
  &=&\frac{(\mu_1^0-a_2)(\mu_2^0-a_3)}{(\mu_1^0-a_3)(\mu_2^0-a_2)}
\exp[(a_2-a_3)(2x-2a_1t+\theta_1^0) ] \ .
\end{eqnarray}
This is a system from which the $\mu_i$'s can be found explicitly.
The solutions
are plotted in Figure~\ref{fig4ab} and the corresponding $U$ and $W$ are
graphed in Figure~\ref{fig4cd}. By initially
choosing different branches of this particular
Riemann surface we see that the solitons `change form', i.e.
a kink changes to
an antikink and vice versa. An analytic explanation is given in the next
section. This was first observed in Alonso and Rues(1992)
who noticed this
phenomenon asymptotically. Using the algebraic geometric construction the
finite-time interactions can be analyzed as well.

\subsection{Change of Form of Kinks to Antikinks.} The solutions obtained from
(\ref{systemofmui}) can be put in the following form
\begin{eqnarray}
  \mu_1&=&\frac{-a_1g_1-a_2g_2-a_3g_3}{g_1+g_2+g_3} \label{mu1cof}\ ,\\
  \mu_2&=&\frac{a_1h_1+a_2h_2+a_3h_3}{h_1+h_2+h_3} \label{mu2cof}\ ,
\end{eqnarray} where
\begin{eqnarray}
  g_1&=&\exp\left[-2a_1x-2a_2a_3t-a_1\theta_2^0+\log\frac{(a_3-a_2)(\mu_1^0-a_
      1)}{\mu_2^0-a_1}\right] \ ,\\
  g_2&=&\exp\left[-2a_2x-2a_1a_3t-a_2\theta_1^0+\log\frac{(a_1-a_3)(\mu_1^0-a_
      2)}{\mu_2^0-a_2}\right] \ ,\\
  g_3&=&\exp\left[-2a_3x-2a_1a_2t-a_3(\theta_1^0+\theta_2^0)+\log\frac{(a_2-a_
      1)(\mu_1^0-a_3)}{\mu_2^0-a_3}\right]
  \ ,\\
  h_1&=&\exp\left[2a_1x+2a_2a_3t+a_1\theta_2^0+\log\frac{(a_3-a_2)(\mu_1^0-a_1
      )}{\mu_2^0-a_1}\right] \ ,\\
  h_2&=&\exp\left[2a_2x+2a_1a_3t+a_2\theta_1^0+\log\frac{(a_1-a_3)(\mu_1^0-a_2
      )}{\mu_2^0-a_2}\right] \ ,\\
  h_3&=&\exp\left[2a_3x+2a_1a_2t+a_3(\theta_1^0+\theta_2^0)+\log\frac{(a_2-a_1
      )(\mu_1^0-a_3)}{\mu_2^0-a_3}\right] \ .
\end{eqnarray}
Notice the similarity between the form of the
solutions (\ref{mu1cof}),(\ref{mu2cof}) and
the solution (\ref{burgerlike}) for $U$ in
the previous case. Since they have an identical form as that in
(\ref{burgerlike}), all the analysis from the last section applies and we
conclude that both $\mu_1$ and $\mu_2$ are kinks which
experience fission or fusion respectively.
We find that as $t\rightarrow-\infty$, $\mu_1$ is an antikink
and $\mu_2$ decomposes into two kinks. This implies that $-\frac{1}{4}U
=\mu_1+\mu_2$
consists of two kinks and one antikink. As $t\rightarrow\infty$, $\mu_1$
fissions into two antikinks and the 2 kinks comprising $\mu_2$
fuse into one kink.
Therefore $\mu_1+\mu_2$
is the sum of one kink and two antikinks. This explains the
transformation of kinks to antikinks and vice versa. See
Figure ~\ref{fig:change} to see how these $\mu$ variables
combine to form $U$. Of
course the same analysis can be performed as above to see this analytically.
This is the first time two-soliton solutions of this equation have been
derived in this simple form.

\section{The SHG Equations}

Solutions of the cKdV may also be transformed into solutions of the SHG
equation if $\kappa$ is chosen to be $-1$. This is shown explicitly in
Appendix \ref{app:A}.
The results from the previous sections can be viewed in the context of
the SHG equations.  First, if $\kappa=-1$, the
initial conditions must be chosen
differently or the $\mu_i$'s will not be real valued.
That is, the cases $\kappa=1$ and $\kappa=-1$ are
dual to each other in the sense that in the process of the
construction
of  the Riemann
surface the cuts in the complex plane for the $\kappa=1$ case correspond to
where real valued solutions lie in the $\kappa=-1$ case and vice versa.
This is because the kappa appears under the square root in (\ref{mx}) so that
real values of $\mu$ for $kappa=1$ correspond exactly to imaginary values of
$\mu$ when $\kappa=-1$ and vice versa.

Another important difference is that no kink solutions exist for the SHG
equations. When $\kappa=-1$ there is no way to deform the Riemann
surface so that all points coalesce piecewise
as required for kink solitons. Therefore there are always at
least two hyperelliptic points left. This means that SHG solutions do not
have a possibility of
either fusing or fissioning and no change of form can occur.

The phase shift formulas for this system are very similar to those
in the cKdV hierarchy with $\kappa=1$, namely
\begin{eqnarray}
  \Delta_1&=&\frac{1}{\sqrt{-(a_2-m_5)(a_2-m_6)}}
  \log\left|\frac{M(a_2)+M(a_1)}{M(a_2)-M(a_1)}\right| \ ,\\
  \Delta_2&=&\frac{1}{\sqrt{-(a_1-m_5)(a_1-m_6)}}
  \log\left|\frac{M(a_1)+M(a_2)}{M(a_1)-M(a_2)}\right| \ .
\end{eqnarray}
This is the first time these formulas have been derived. It is remarkable
that the formulas are so similar to those of the coupled KdV equation since
they are not in the same hierarchy of equations - the two equations are
derived from two different potentials. The phase shift formulas show
another strength of the algebraic geometric method since the details of the
inverse scattering transform have not been completed at this time, see
Khusnutdinova(1998).
$u$ and
$v$ are plotted in Figure ~\ref{fig:shg}.  These functions can be transformed
into $q_1,q_2$ of the SHG by transformations in Appendix \ref{app:A}.

\section{Modified Coupled Dym Equations}

\subsection{Genus-1 solutions for the cDym System}
\subsubsection{Periodic solutions.} A periodic  solution of the cDym equation
is described by the following differential equation
\begin{equation}\label{percd}
  \frac{\sqrt{\mu} d\mu}{\sqrt{\prod_{i=1}^4 (\mu-m_i)}} = dX\ ,
\end{equation} for particular
choice of $m_i$'s. Here $X=2x-2ct+\theta_0$ and integration is carried
out on the Riemann surface $W^2={\displaystyle \frac{C(E)}{E}}$.
This can be reduced to a standard form by
introducing a new variable $Y$ by
\begin{equation}\label{xx1}
 dY = \frac{dX}{\sqrt{\mu}} \ .
\end{equation}
After integration (\ref{percd}) becomes
\begin{equation}\label{intdmu}
  \int_{\mu_0}^{\mu} \frac{ d\mu}{\sqrt{\prod_{i=1}^4 (\mu-m_i)}} =
  Y \ .
\end{equation}
Notice that this holomorphic differential is defined on a genus two Riemann
surface. To invert this integral one has first to consider the following
problem of inversion
\begin{eqnarray}
  \int_{\mu_1^0}^{\mu_1} \frac{d\mu_1}{\sqrt{\mu_1 C_4 (\mu_1)}} +
  \int_{\mu_2^0}^{\mu_2}
  \frac{d\mu_2}{\sqrt{\mu_2 C_4 (\mu_2)}} &=& \theta_1^0\ , \label{pinv1}\\
  \int_{\mu_1^0}^{\mu_1} \frac{\mu_1 d\mu_1}{\sqrt{ \mu_1 C_4 (\mu_1)}} +
  \int_{\mu_2^0}^{\mu_2} \frac{\mu_2 d\mu_2}{\sqrt{ \mu_2 C_4 (\mu_2)}} &=&
  X_1 + \theta_2^0\ , \label{pinv2}
\end{eqnarray}
where $C_4(\mu) =  \prod_{l=1}^4 (\mu - m_j)$.
One needs to rearrange integrals in such a way that to obtain
$X_1$ on the right hand side of the first integral equation.  This yields
exact formulas for $\mu_1$ and $\mu_2$ in terms of Riemann
$\theta$-functions. By fixing
$\mu_2=m_3$ and writing $\mu=\mu_1$ we resolve the initial problem
of inversion. (For details see Alber and Fedorov(1999).)

\subsubsection{Kink solutions.}
Now consider a kink limit by setting $m_1,m_2 \rightarrow a_1$ and $m_3,m_4
\rightarrow a_2$ such that $a_2 > a_1 > 0$. Integral (\ref{intdmu}) becomes
\begin{equation}\label{kinkcd1}
  \frac{ d\mu}{(\mu-a_1)(\mu-a_2)} = dY\ .
\end{equation}
Observe that this is the same problem of inversion as in the cKdV case except
that we have $Y$ instead of $X$ on the right hand side.
After integrating we obtain
\begin{equation}\label{kinkfr}
  \mu(Y)= \frac{a_1(\mu_0-a_2)  -a_2(\mu_0 -a_1) \exp ((a_1 -a_2)Y)}
{(\mu_0-a_2) - (\mu_0-a_1) \exp ((a_1-a_2)Y)}\ .
\end{equation}
This gives $\mu$ as a function of $Y$ and as in the cKdV case this is a
kink.  $X$ is defined in terms of $Y$
by  (\ref{xx1}) once we know $\mu(Y)$ from (\ref{kinkfr}). The
integration in (\ref{xx1}) may be carried out explicitly.
Note that ${dX}/{dY}>0$ and so $X(Y)$ is always increasing.  By the
definition of $X$ it is also clear that the range of $X(Y)$ is all
real numbers.  Therefore the
Inverse Function Theorem implies that an inverse function $Y=Y(X)$ exists
for all values of $X$ and is monotonically decreasing.
Therefore the graph of $\mu(X)=\mu(Y(X))$ will have a similar appearance
to that of $\mu(Y)$, that is, it
is also a kink.

After combining numerics for $Y(X)$ with the expression for $\mu$, a
description for the kink of the cDym system is obtained, see
Figure~\ref{fig:cdymkink}.

\subsubsection{Cusp solution}
For this solution, the limit $m_1,m_2 \rightarrow a_1$ and $m_3,m_4
\rightarrow a_2$ such that $a_2< 0< a_1$ is analyzed.
The analysis is the same as in
the kink case except that now ${dX}/{dY}$ changes sign exactly once
when the branch point $Y^*$ is crossed where $\mu(Y^*)=0$.  In this case,
$Y(X)$ has two branches begining at the hyperelliptic point $X^*$ where
$Y(X^*)=Y^*$.  Therefore $\mu(X)$ has two branches and reaches a cusp at
the point $X^*$.  See Figure~\ref{fig:cdymcusp}.

\subsubsection{Peakon  solution }
If the Camassa-Holm shallow water equation is an
indication(1994), a peakon may develop
in the limit $m_1,m_2 \rightarrow a_1$ and $m_3,m_4
\rightarrow a_2$ with $a_2= 0< a_1$. The analysis is similar to that of
the kink case, except now the range of $X(Y)$ is bounded above by
some number $X^*$. This means that the inverse function $Y(X)$ is defined
only for those $X<X^*$.  But, it can be defined symmetrically, as if
the integration were carried out on the negative branch of the square
root, and this gives rise to a peakon solution.  The difference between
this and a cusp solution is that in the cusp,
\begin{equation}
 \frac{dY}{dX}=\frac{1}{\sqrt{\mu}} \ ,
\end{equation}
is infinite at the branch point, while in the Peakon case $\mu(Y)>0$ for all
$Y$ and so at the branch point the derivative is finite.

\subsection{Genus 2 solutions}
The algebraic geometric procedure outlined thus far in this paper can be
used for other equations as well, even when other methods may
fail. (For details see
Alber and Fedorov (1999).) Using
our experience with the cKdV system, the case when there are
three double points will be considered.
For the positive branch of the Riemann Surface
$W^2=C(E)/E$, the problem of inversion can be written as
\begin{eqnarray}
  \frac{\mu_1 d\mu_1}{2(\mu_1-a_1)(\mu_1-a_2)(\mu_1-a_3)\sqrt{\mu_1}}+
  \frac{\mu_2
    d\mu_2}{2(\mu_2-a_1)(\mu_2-a_2)(\mu_2-a_3)\sqrt{\mu_2}} &=& dt \ ,\\
  \frac{\mu_1^2 d\mu_1}{2(\mu_1-a_1)(\mu_1-a_2)(\mu_1-a_3)\sqrt{\mu_1}}+
  \frac{\mu_2^2 d\mu_2}{2(\mu_2-a_1)(\mu_2-a_2)(\mu_2-a_3)\sqrt{\mu_2}} &=&
  dx \ .
\end{eqnarray}
This inversion problem is very similar to that in the cKdV case, with the
exception of the poles present in the left hand side of the equation. This
case is complicated by the fact that only differentials of the third kind
appear having simple poles at $a_1,a_2,a_3$. Following the procedure
outlined in Alber and Fedorov(1999),
a third variable $y$ is introduced such
that
\begin{equation}
  \frac{ d\mu_1}{2(\mu_1-a_1)(\mu_1-a_2)(\mu_1-a_3)\sqrt{\mu_1}}+ \frac{
    d\mu_2}{2(\mu_2-a_1)(\mu_2-a_2)(\mu_2-a_3)\sqrt{\mu_2}} = dy \ .
\end{equation}
Also   the normalized differentials of the third kind are introduced
\begin{equation}
  \Omega_i=\frac{\alpha_i d\mu}{(\mu-\alpha_i^2)\sqrt{\mu}}\
,\qquad i=1,2,3 \ ,
\end{equation}
where $\alpha_i=\sqrt{a_i}$. Next, consider three points $z_i$ given by
\begin{equation}\label{system}
  \sum_{j=1}^3 \int_{P_0}^{P_j}\Omega_i=z_i \ ,\qquad i=1,2,3\ .
\end{equation}
The $z_i$ are dependent on $x,t,y$ by
\begin{eqnarray}\label{transform}
  z_1&=&2\sqrt{a_1}\,[x-(a_2+a_3)t+a_2a_3y] \ ,\\
  z_2&=&2\sqrt{a_2}\,[x-(a_1+a_3)t+a_1a_3y] \ ,\\
  z_3&=&2\sqrt{a_3}\,[x-(a_1+a_2)t+a_1a_2y] \ .
\end{eqnarray}
Integrating (\ref{system}) and putting $\xi=\sqrt{\mu}$ gives
\begin{equation}
  \frac{(\xi_1-\alpha_i)(\xi_2-\alpha_i)(\xi_3-\alpha_i)}{(\xi_1+\alpha_i)(\xi
    _2+\alpha_i)(\xi_3+\alpha_i)} = e^{z_i} \ ,\qquad i=1,2,3\ .
\end{equation}
This gives the following system for the symmetric polynomials in $\xi$.
\begin{eqnarray}
\left( \begin{array}{ccc} 1-e^{z_1} & -\alpha_1(1+e^{z_1}) &
      \alpha_1^2(1-e^{z_1}) \\
      1-e^{z_2} & -\alpha_2(1+e^{z_2}) & \alpha_2^2(1-e^{z_2}) \\ 1-e^{z_3}
      & -\alpha_3(1+e^{z_3}) & \alpha_3^2(1-e^{z_3}) \end{array} \right)
\left(
\begin{array}{c}
{\xi_1\xi_2\xi_3} \\ {\xi_1\xi_2+\xi_2\xi_3+\xi_1\xi_3} \\
{\xi_1+\xi_2+\xi_3} \end{array} \right)
=\left(
\begin{array}{l}{\alpha_1^3(1+e^{z_1})} \\
{\alpha_2^3(1+e^{z_2})} \\
{\alpha_3^3(1+e^{z_3})}
\end{array}
\right)
\end{eqnarray}
>From this, the expression $\mu_1+\mu_2+\mu_3=\xi_1^2+\xi_2^2+\xi_3^2$
may be found as a
function of $z_1,z_2,z_3$.
The determinant of the matrix in the system must be zero to obtain
nonzero solutions. This
added equation gives $\mu_3$ in terms of $\mu_1$ and $\mu_2$.
Then we can find $\mu_1+\mu_2$ as a function of $z_1,z_2$. Then this
solution must be connected to the $(x,t)$ variables by using
(\ref{transform}).

Numerics are provided in
Figures~\ref{fig:dymfission} and \ref{fig:dymchange}.
As expected, the phenomena of change of form and fission occurs.
A more detailed analysis of fission/fusion for the cDym case will be described
in a forth coming paper.

\section{Acknowledgments}
The research of Mark Alber and Gregory Luther was partially supported
by NSF grant DMS 9626672.
Mark Alber would like to thank Peter Miller for a helpful
discussion and bringing to his attention references Estevez
{\it et~al.}(1994)
and Martinez Alonso and  Medina Reus(1992).

\appendix
\section{Transformations to Related Equations}
\label{app:A}
In what follows we describe exact connections between solutions of the cKdV
and the Kaup equations, Boussinesq systems and SHG system.

\subsection{Generalized Coupled KdV System} For this system,
assume that $\kappa =1$ and $r=0$. Observe that if we take $n=1$, $B$ takes
the form of $B=E+b_1$.  Furthermore, as in (\ref{K}) we see that $b_1=-u/2
+K_1$.  Collecting coefficients of order 1 and 0 respectively in
(\ref{gneq}) and substituting the value of $b_1$ shows that $u,v$ satisfy
the modified cKdV system
\begin{eqnarray}
  u_t &=& v' -{\textstyle\frac{3}{2}} uu' + K_1 u' \ , \label{cKdV1}\\
  v_t &=&{\textstyle\frac{1}{4}}u''' -vu' -{\textstyle\frac{1}{2}}uv' +K_1
  v' \ . \label{cKdV2}
\end{eqnarray}
$K_1$ is a small parameter. When $K_1=0$ this is the cKdV.

\subsection{Generalized Boussinesq System} To connect $u$ and $v$ to the
classical Boussinesq system the following change of variables must be made
\begin{eqnarray}
  u(x,t) &=& -{\textstyle\frac{1}{2}}U(X,T) \ ,\label{coupledKdV1}\\
  v(x,t) &=& {\textstyle\frac{1}{16}}U(X,T)^2
  -{\textstyle\frac{1}{4}}W(X,T) \ ,\label{coupledKdV2}
\end{eqnarray}
where $X = x$ and $T = -t/2$.  Plugging in $u$ and $v$ from
(\ref{coupledKdV1})-(\ref{coupledKdV2}) into (\ref{cKdV1})-(\ref{cKdV2}) and
defining $\gamma = -2K_1$ we see that $U$ and $W$ satisfy
\begin{eqnarray}
  U_T +W_X +UU_X &=& \gamma U_X \ ,\label{bs1}\\
  W_T +U_{XXX}+(WU)_X &=& \gamma W_X \ .\label{bs2}
\end{eqnarray} Again
$\gamma$ is a small parameter. When $\gamma =0$, we have exactly the
Boussinesq System. Otherwise the system gives rise to a generalization of
the Boussinesq system. Such a transformation can be found in
Sattinger(1995) for example.

\subsection{Generalized Kaup Equations} To connect $U(X,T)$ and $W(X,T)$
from above to $\pi(x,t)$ and $\phi(x,t)$ from the Kaup equations
(\ref{swka}),(\ref{swkb}) the following change of variables was suggested by
Kaup(1972):
\begin{eqnarray}
U(X,T)&=&\frac{\epsilon}{\beta}\phi_x(x,t) \ ,\\
W(X,T)&=&\beta^{-2}[1-\epsilon\pi(x,t)] \ ,
\end{eqnarray}
where $X=x$, $T=\beta t$, $\beta=\delta\sqrt{3}/\sqrt{1 - 3\sigma}$, and
$\alpha=\beta \gamma$.  Then substituting this into (\ref{bs1})-(\ref{bs2})
gives
\begin{eqnarray}
  \pi_t &=&\phi_{xx} +{\textstyle\frac{1}{3}} (1 -
  3\sigma)\delta^2\phi_{xxxx} -\epsilon (\phi_x\pi)_x +\alpha \pi_x \ ,\\
  \pi &=&\phi_t +{\textstyle\frac{1}{2}}\epsilon\phi_x^2 -\alpha \phi_x \ ,
\end{eqnarray}
which is a perturbed Kaup equations and reduces identically to it
when $\alpha=0$.  Summarizing we see that every solution of the coupled KdV
system yields, using the transformations stated, a solution of the
Boussinesq and Kaup systems.

\subsection{The SHG System}
For the SHG system we choose $\kappa=-1$ in
the potential (\ref{v}), and define $w,\nu$ by the
equations $u=w$, and $v=\nu_x/2+\nu^2/4$. Then the generating equation
(\ref{eq}) becomes

\begin{eqnarray}\label{aone}
  Ew_t+{\textstyle\frac{1}{2}}\nu_{xt}+{\textstyle\frac{1}{2}}\nu\nu_t &=&
  -{\textstyle\frac{1}{2}}B_{xxx} + 2B_xwE + B_x\nu_x +
  {\textstyle\frac{1}{2}}B_x\nu^2 \nonumber\\
  &&-2B_xE^2 +Bw_xE+{\textstyle\frac{1}{2}} B \nu_{xx} +
  {\textstyle\frac{1}{2}} B\nu\nu_x \ .
\end{eqnarray} Now we choose $B={b}/{E}$, that is we choose
$m=1,n=-1,b=b_{-2}$. Then (\ref{aone}) becomes
\begin{eqnarray}
  Ew_t+{\textstyle\frac{1}{2}}\nu_{xt} + {\textstyle\frac{1}{2}}\nu\nu_t &=&
  -{\textstyle\frac{1}{2E}} b_{xxx} + 2b_xw + {\textstyle\frac{1}{E}}
  b_x\nu_x
  + {\textstyle\frac{1}{2E}} b_x\nu^2\nonumber\\
  && - 2b_xE + bw_x + {\textstyle\frac{1}{2E}} b\nu_{xx} +
  {\textstyle\frac{1}{2E}} b\nu\nu_x\ .
\end{eqnarray}
Collecting coefficients
of $E$ gives
\begin{equation}
  b=-\frac{\eta}{2}\ ,
\end{equation}
where we define $\eta$ by $\eta_x=w_t$. Then collecting
coefficients of orders zero and
one respectively and substituting in the value for $b$ gives
\begin{eqnarray}\label{shg}
{\textstyle\frac{1}{2}}\nu_{xt}+{\textstyle\frac{1}{2}}\nu\nu_t &=&
-\eta_xw- {\textstyle\frac{1}{2}}\eta w_x \,\\
\label{shg2}
0&=&{\textstyle\frac{1}{4}}\eta_{xxx}-{\textstyle\frac{1}{2}}\eta_x\nu_x
-{\textstyle\frac{1}{4}}\eta_x\nu^2-{\textstyle\frac{1}{4}}\eta\nu_{xx}-
{\textstyle\frac{1}{4}}\eta\nu\nu_x \ .
\end{eqnarray}
(\ref{shg2}) may be integrated to get
\begin{equation}
(\eta_x)_x= (\eta\nu)\nu+\eta\nu_x+\int\nu_x(\eta\nu-\eta_x) dx \ .
\end{equation}
Notice that this equation is satisfied when $\eta_x=\eta\nu$. Next we define
the new function $s$ by the relation $\nu_t=s^{-1}-\eta w$. Then plugging
this into (\ref{shg}) gives
\begin{equation}
-\frac{s_x}{s^2}-\eta_xw-\eta w_x + \frac{\nu}{s} - {\eta\nu w} =
-2\eta_xw- \eta w_x\ .
\end{equation}
Substituting $\eta_x=\eta\nu$ and canceling like terms yields $s_x=s\nu$.
These two equations, along with the two we defined will determine the SHG
system.  Summarizing, we have
\begin{eqnarray}\label{foureq}
\eta_x&=&\eta\nu=w_t \ ,\\
\nu_t&=&s^{-1}-\eta w \ ,\\ s_x&=&s\nu \ .
\end{eqnarray} Next define $Q=s^{-1}$ and $\phi_t =\eta$. This and
$\eta_x=w_t$ implies that $w=\phi_x$. Therefore (\ref{foureq}) becomes
\begin{eqnarray} \label{qphi} (Q\phi_t)_x &=& Q_x\phi_t+Q \phi_{xt}
  \nonumber \ ,\\ &=& -\frac{s_x}{s^2}\eta+\frac{\eta_x}{s} \nonumber \ ,\\
  &=& \frac{1}{s}(\eta_x-\eta\nu)=0 \ ,
\end{eqnarray} and
\begin{eqnarray}\label{q}
(\ln Q)_{xt}-\phi_x\phi_t &=& -(\ln s)_{xt}-\eta w \nonumber \ ,\\
&=& -\left(\frac{s_x}{s}\right)_t -\eta w \nonumber \ ,\\
&=& -\nu_t -\eta w \nonumber \ ,\\
&=& -(s^{-1}-\eta w)-\eta w =-Q \ .
\end{eqnarray}
Finally, substitute $q_1=(\sqrt{Q}/2)\exp[i(\phi/2)]$ and the real and
imaginary parts of the following equation correspond to
(\ref{qphi}) and (\ref{q}), so that
\begin{equation}
q_{1xt}q_1^*-q_{1t}^* q_{1x}=-2(q_1 q_1^*)^2\ .
\end{equation}
If we define $q_2=-[(\ln Q)_x +i\phi_x]\exp(i\phi)/4$. Then $q_1$ and $q_2$
satisfy
\begin{eqnarray}
q_{1x}&=&-2q_2q_1^* \ ,\\
q_{2t}&=&q_1^2 \ ,
\end{eqnarray}
which is exactly the SHG equation, and
$q_1$ and $q_2$ are obtained from the $\mu$ variables through the relation
\begin{eqnarray}
\phi_x&=&u \ ,\\
Q &=& \exp\left(-\frac{u_t}{\int{u_t \;dx}}\right) \ .
\end{eqnarray}

The above transformations were inspired by Khusnutdinova(1998).
Summarizing the above gives that
solutions of the SHG system may be expressed as follows
\begin{eqnarray}
  q_1 &=& \frac{1}{2} \exp \left ( \frac{1}{2} \int u_t
    \left(\int{u_t\;dx}\right)^{-1} \; dx \right ) \; \exp \left(\frac{i}{2}
    \int u \; dx \right) \ ,\label{q12a}\\ q_2 &=& \frac{1}{4}\left(
     \frac{d^2}{dx^2} \log \left(\int{u_tdx}\right)
     -iu\right) \; \exp \left(i \int u \; dx
  \right) \ .\label{q12b}
\end{eqnarray}
Notice that the dependence on $v$ is implicit
by the fact that $\nu=u_t \left(\int{u_t\;dx}\right)^{-1}$.

\section{Trace Formulas}
\label{app:B}

\subsection{cKdV System}
In this section,
the connection between $u,v,$ and $\mu$ for cKdV is derived. This
connection is
called the trace formula for this system. Collect
coefficients of $E^{n-r+2}$ in (\ref{eq}) to see that $2\kappa
b_0^\prime=0$ from which we assume that $b_0 =1$. Now gather coefficients of
order $n-r+1$ to arrive at
\begin{equation}
  \label{K} b_1 = -\frac{u}{2\kappa} +K_1 \ ,
\end{equation}
for some constant of integration
$K_1$. Next collect coefficients of order $2n-2r+1$ in
(\ref{seq}) to see, along with (\ref{K}), that
\begin{equation}
  K_1  =-\frac{1}{2} \sum_{i=1}^{2n+2}m_i \ .
\end{equation}
(\ref{K}) then  yields the trace formula for $u$,
\begin{equation}
u=2\kappa \sum_{i=1}^{n} \mu_i -\kappa\sum_{i=1}^{2n+2}m_i \ .
\end{equation}

Now we just need the trace formula for $v$. For this we collect coefficients
of order $n-r$ in (\ref{eq}) to get
\begin{equation}
  0=2\kappa b_2^\prime+2b_1^\prime u+b_1u^\prime+v^\prime \ ,
\end{equation}
which gives upon substitution of $b_1$ from (\ref{K}) that
\begin{equation}
  v^\prime=-2\kappa b_2^\prime +\frac{3uu^\prime}{2\kappa }-K_1 u^\prime \ ,
\end{equation}
or simply
\begin{equation}
  \label{K2} v=-2\kappa b_2 +\frac{3u^2}{4\kappa} -K_1 u+K_2 \ ,
\end{equation}
where $K_2$ is some constant and $K_1$ is the same as above. From the
definition of $B(E)$ we see that
\begin{equation}
b_2 = \sum_{1\le i<j\le n} \mu_i \mu_j \ .
\end{equation}
All that remains is to find $K_2$. To derive this we collect coefficients of
order $2n-2r$ in (\ref{seq}). To simplify calculations let $c_{n}$ be the
$(2n-2r)^{th}$ coefficient of the polynomial $C(E)$. Then we see that
\begin{equation}
  2\kappa(2b_2 +b_1^2)+2u(2b_1)+2v=c_{n} \ ,
\end{equation} or after substitution of the
value of $b_1$ that
\begin{equation}
\label{c2n} 4\kappa b_2 -\frac{3u^2}{2\kappa} +2uK_1 +2\kappa K_1^2
+2v=c_{n} \ .
\end{equation}
Now solving for $K_2$ in (\ref{K2}) and substituting in $v$ from
(\ref{c2n}),
\begin{eqnarray}
  K_2&=&v +2\kappa b_2 - \frac{3u^2}{4\kappa} +K_1u \ ,\\
  &=&\left(\frac{c_{2}}{2} -2\kappa b_2 +\frac{3u^2}{4\kappa} -uK_1 -\kappa
  K_1^2\right)+2\kappa b_2 -\frac{3u^2}{4\kappa} + K_1u \ ,\\
  &=& \frac{c_{2}}{2} -\kappa K_1^2 \ .
\end{eqnarray}
Now from the definition of $C(E)$ we see that
\begin{equation}
c_{n}=2\kappa \sum_{1\le i<j\le 2n+2} m_i m_j \ .
\end{equation}
\subsection{cDym System}
In an analogous manner, we derive the trace formulas for the cDym system.
The derivation of the
trace formula for $u$ is identical to the cKdV case so that
\begin{equation}
u=2 \sum_{i=1}^{n} \mu_i -\sum_{i=1}^{2n+2}m_i \ .
\end{equation}

Only the trace formula for $v$ is different.  For this we collect
coefficients of order $n-r$
in (\ref{eqdym}) to get
\begin{equation}
  0=-\frac{b_1'''}{2}+2b_2^\prime+2b_1^\prime u+b_1u^\prime+v^\prime \ ,
\end{equation}
which, upon substitution of $b_1$ from (\ref{K}) and integrating,
gives
\begin{equation}
\label{K2dym} v=-\frac{u''}{4}-2 b_2 +\frac{3u^2}{4} -K_1 u+K_2 \ ,
\end{equation}
where $K_2$ is some constant and $K_1$ is the same as in cKdV. From the
definition of $B(E)$ we see that
\begin{equation} b_2 = \sum_{1\le i<j\le n} \mu_i \mu_j \ . \end{equation}
All that remains is to find $K_2$. To derive this we collect coefficients of
order $2n-2r-1$ in (\ref{seqdym}). To simplify calculations let $c_{n-1}$ be
the $(2n-2r-1)^{st}$ coefficient of the polynomial $C(E)$. Then we see that
\begin{equation}
-b_1''+2(2b_2 +b_1^2)+2u(2b_1)+2v=c_{n-1} \ ,
\end{equation}
or after substitution of the value of $b_1$ that
\begin{equation}\label{c2ndym}
\frac{u''}{2}+4b_2 -\frac{3u^2}{2} +2uK_1 +2 K_1^2 +2v=c_{n-1} \ .
\end{equation}
Now solving for $K_2$ in (\ref{K2dym}) and substituting in $v$ from
(\ref{c2ndym}) gives
\begin{eqnarray}
  K_2&=&v +2 b_2 - \frac{3u^2}{4} +K_1u +\frac{u''}{4}\ ,\\
  &=&(\frac{c_{2}}{2} -2 b_2 +\frac{3u^2 }{4}-uK_1 - K_1^2-\frac{u''}{4})+2
  b_2 -
  \frac{3u^2}{4} + K_1u +\frac{u''}{4}\ ,\\
  &=& \frac{c_{n-1}}{2} - K_1^2 \ .
\end{eqnarray}
>From the definition of $C(E)$ we see that
\begin{equation}
c_{n-1}=2\sum_{1\le i<j\le 2n+2} m_i m_j \ .
\end{equation}

\newpage
\section*{Bibliography}
\addcontentsline{toc}{section}{Bibliography}
\begin{description}

\item
D. Mumford, {\it Tata Lectures on Theta I and II, Progress
in Math.28 and 43,} (Birkhauser, Boston 1983).

\item
Ercolani, N. and H. McKean, ``Geometry of KdV(4). Abel
sums, Jacobi variety, and theta function in the scattering case,''
Invent. Math {\bf 99,} 483-544 (1990).

\item  Ablowitz, M.J. and Y-C. Ma,  ``The Periodic Cubic
Schr\H{o}dinger Equation,''
Studies in Appl. Math. {\bf 65,} 113--158 (1981).

\item  Alber, M.S. and S.J. Alber, `` Hamiltonian formalism for
finite-zone solutions
of integrable equations,'' C. R. Acad. Sci. Paris Sr. I Math. {\bf
301,} 777--781 (1985).

\item  Alber, M.S. and J.E. Marsden,
``On Geometric Phases for Soliton
Equations,'' Comm. Math. Phys. {\bf 149,} 217--240 (1992).

\item
Antonowicz, M. and A.P. Fordy,  ``A family of completely
integrable multi-Hamiltonian systems'',
Phys. Lett. A {\bf 122,} 95--99 (1987).

\item
Antonowicz, M. and A.P. Fordy, ``Coupled KdV equations with
multi-Hamiltonian structures,'' Physica D {\bf 28,} 345--357 (1987).

\item
Antonowicz, M. and A.P. Fordy, `` Coupled Harry Dym equations with
multi-Hamiltonian structures,'' J. Phys. A {\bf 21,} L269--L275 (1988).

\item
Antonowicz, M. and A.P. Fordy, `` Factorization of energy dependent
Schr\H{o}dinger operators: Miura maps and modified systems,''
Comm. Math. Phys. {\bf 124,} 465--486 (1989).

\item
Alber, M.S., G.G. Luther, and J.E. Marsden, `` Energy Dependent
Schr\H{o}dinger
Operators and Complex Hamiltonian Systems on Riemann Surfaces,''
{\it Nonlinearity\/} {\bf 10,} 223--242 (1997).

\item
Channell, P.J. and C. Scovel, ``Symplectic integration
of Hamiltonian systems,'' Nonlinearity {\bf 3,} 231--259 (1990).

\item
Channell, P.J. and C. Scovel, ``An introduction to
symplectic integrators,'' Fields Institute Communications {\bf 10,} 45--58
(1996).

\item  Alber, M.S., R. Camassa, D.D. Holm and J.E. Marsden,
``The geometry of peaked solitons and billiard solutions of a class of
integrable PDE's,'' Lett. Math. Phys. {\bf 32,} 137--151 (1994).

\item  Alber, M.S., R. Camassa, D.D. Holm and J.E. Marsden,
``On Umbilic Geodesics and Soliton Solutions of Nonlinear PDE's,''
Proc. Roy. Soc. London Ser. A {\bf 450,} 677--692 (1995).

\item  Alber, M.S., R. Camassa, Yu.N. Fedorov, D.D.
Holm and J.E. Marsden,
``The geometry of new classes of weak billiard solutions of nonlinear
PDE's,'' (preprint) (1999).

\item  Alber, M.S. and  Yu.N. Fedorov,
``Algebraic Geometric Solutions for Nonlinear Evolution
Equations and Flows on the Nonlinear Subvarieties of Jacobians,''
(preprint) (1999)

\item
Whitham, G.B., {\it Linear and Nonlinear Waves,} (Pure and applied
mathematics,
John Wiley \& Sons, Inc. 1974).

\item
Jaulent, M.,
``On an inverse scattering problem with an energy dependent
potential,'' Ann. Inst. H. Poincare A {\bf 17,} 363--372 (1972).

\item
Jaulent, M. and C. Jean, ``The inverse problem for the
one-dimensional Sch\H{o}dinger operator with an energy dependent potential,''
Ann. Inst. H. Poincare A I, II{\bf 25,} 105--118, 119--137 (1976).

\item
Kaup, D.J., ``A Higher-Order Water-Wave Equation and the Method
for Solving It,''
Prog. Theor. Phys. {\bf 54,} 72--78, 396--408 (1975).

\item
Matveev, V.B. and M.I. Yavor,  ``Solutions presque periodiques et a
$N$-solitons de l\'equation hydrodynamique non lineaire de Kaup,''
Ann. Inst. Henri
Poincare: Sec. A {\bf 31,} 25--41 (1979).

\item
Sachs, R.L., ``On the integrable variant of the Boussinesq
system: Painlev\'e
property, rational solutions, a related many-body system, and equivalence
with the AKNS
hierarchy,'' Physica D {\bf 30,} 1--27 (1998).

\item
Martinez Alonso, L. and E. Medina Reus, ``Soliton interaction
with change form
in the classical Boussinesq system,'' Phys. Lett. A {\bf 167,}
370--376 (1992).

\item
Estevez, P.G., P.R. Gordoa, L. Martinez Alonso,  and E. Medina Reus
, ``On the
characterization of a new soliton sector in the classical Boussinesq
system,'' Inverse Problems {\bf 10,} L23--L27 (1994).

\item
Khusnutdinova, K.R. and H. Steudel,  ``Second harmonic
generation: Hamiltonian
structures and particular solutions,'' J. Math. Phys. {\bf 39,}
3754--3764 (1998).

\item
Sattinger, David and Szmigielski, Jacek, ``Energy dependent scattering
theory,'' Differential and Integral Equations {\bf 8,} 945--959 (1995).

\item
Sattinger, David and Szmigielski, Jacek, ``A Riemann Hilbert problem
for an energy dependent Schr\H{o}dinger operator,'' Inverse Problems
{\bf 12,} 1003--1025 (1996).

\item
 Ablowitz, M.J. and Segur, H. ,
{\it Solitons and the Inverse Scattering Transform,}
(SIAM, Philadelphia, 1981).

\item
Belokolos, E.D., A.I. Bobenko, V.Z. Enol'sii, A.R. Its, and  V.B.
Matveev,
{\it Algebro-Geometric Approach to Nonlinear Integrable Equations,\/}
(Springer-Verlag series in Nonlinear Dynamics, 1994).

\item
Ercolani, N. ``Generalized Theta functions and homoclinic
varieties,'' Proc. Symp. Pure Appl. Math. {\bf 49,} 87--100 (1989).

\item
Kupershmidt, B.A., ``Mathematics of Dispersive Water Waves,'' Commun. Math.
Phys. {\bf 99,} 51--73 (1985).

\end{description}

\newpage
\section*{Figures}

\begin{figure}
\begin{center}
\includegraphics[height=2.5in]{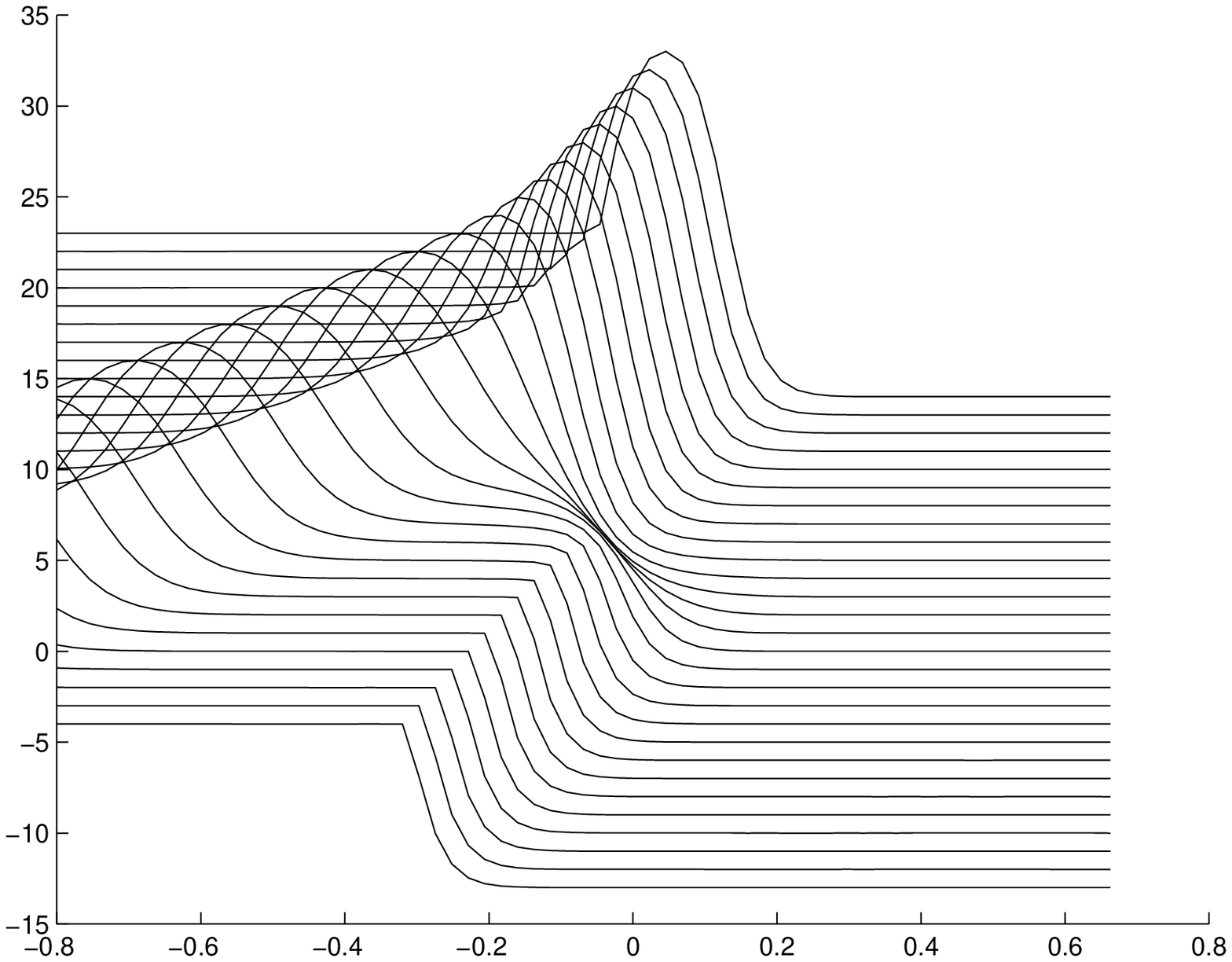}~\includegraphics[height=2.5in]
{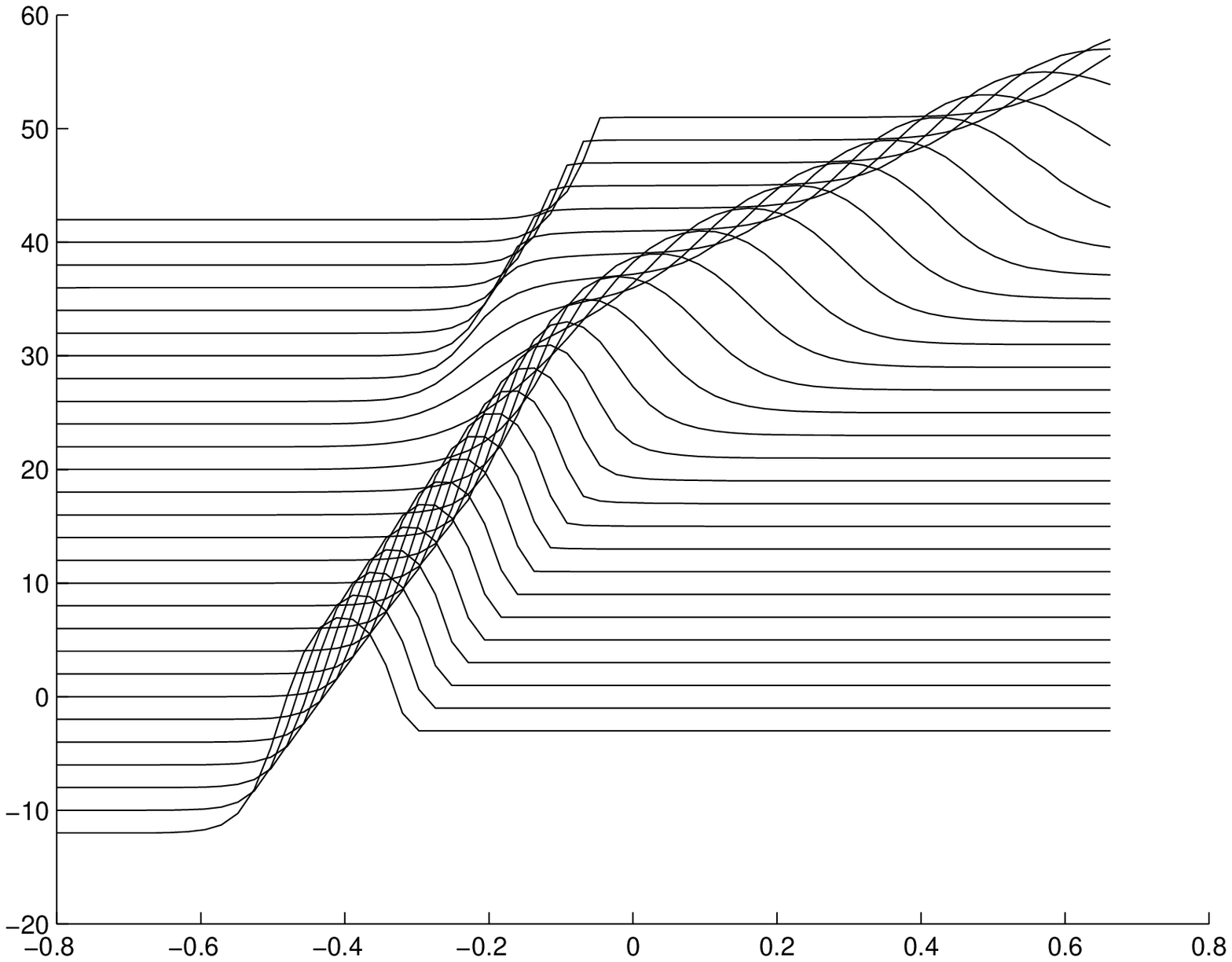}
\end{center}
\caption{\label{fig1cd}These graphs show
$\mu_1$ and $\mu_2$, genus-two solutions with two double roots, for $a1=-5$,
$a2=-14$,$m_5=5$,$m_6=7.5$,$t=-.07 $.  This illustrates two
Kaup type solitons.}
\end{figure}

\begin{figure}
\begin{center}
\includegraphics[height=2.5in]{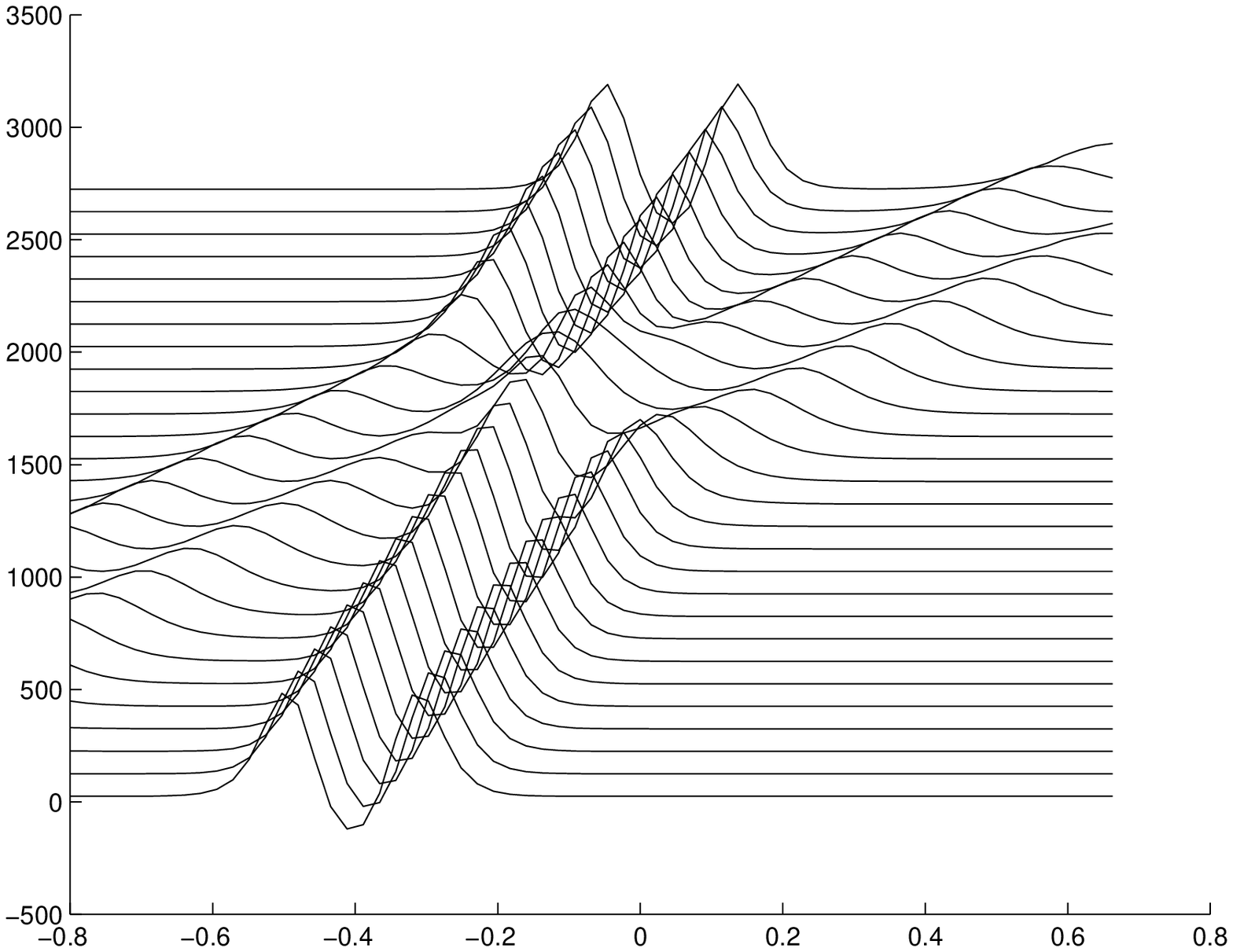}~\includegraphics[height=2.5in]
{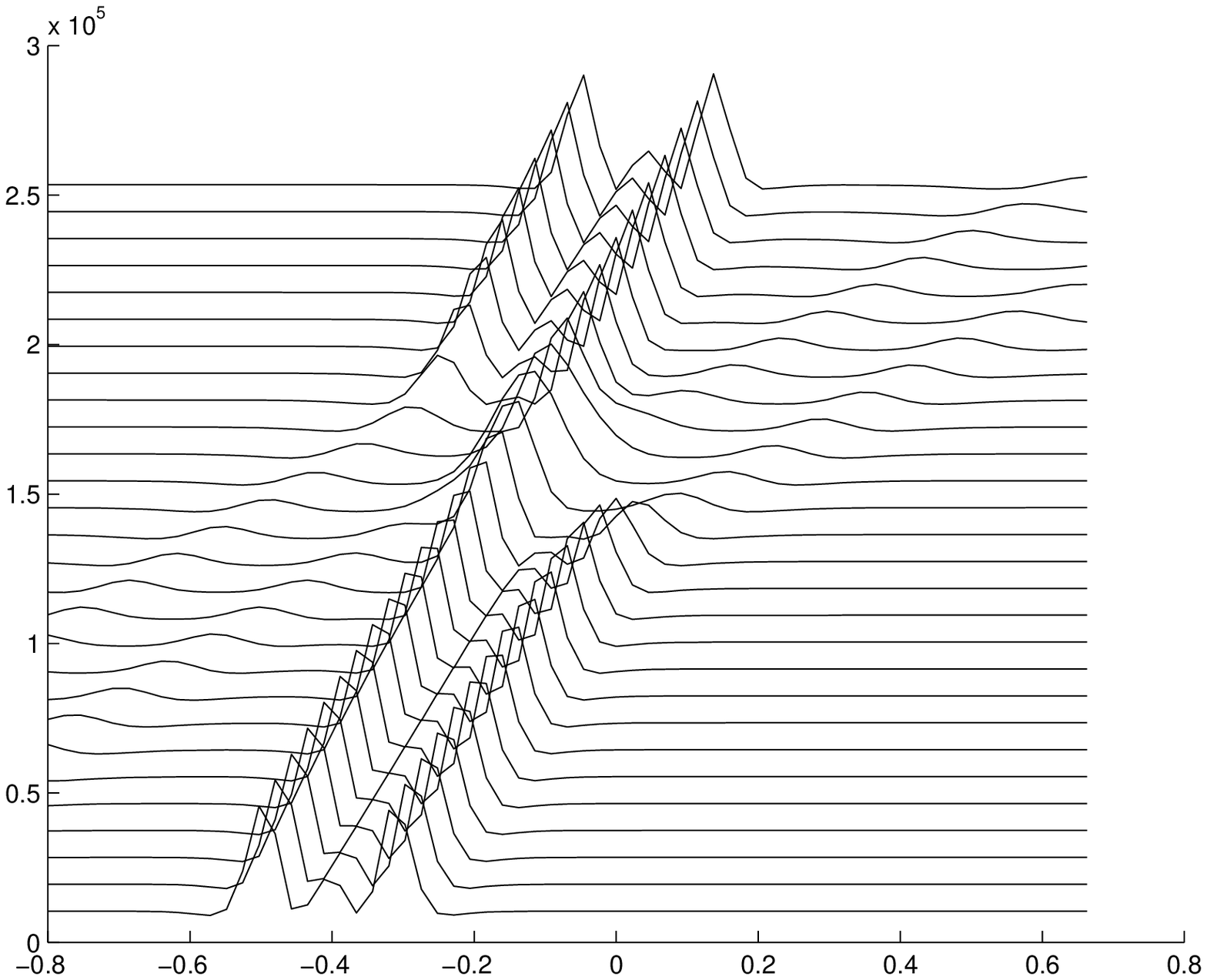}
\end{center}
\caption{\label{fig1ab}This figure shows genus-two solutions
  $U$ and $W$ to the Boussinesq equation derived from (\ref{an1}),
  (\ref{an1part2}) and
  (\ref{coupledKdV1}), (\ref{coupledKdV2}).  Here  $a1=-5$,
  $a2=-14$,$m_5=5$,$m_6=7.5$,$t=-.07 \dots .07$.}
\end{figure}

\begin{figure}
\begin{center}
\includegraphics[height=3in]{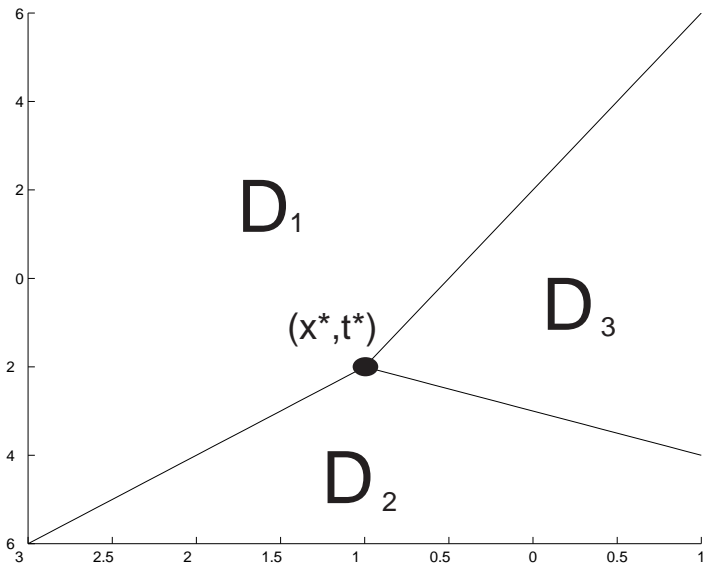} \end{center}
\caption{\label{fusion}Areas $D_1,D_2,D_3$ where away from the boundaries,
  $U$ from the Boussinesq equation has a constant height
  of $a_1,a_2,a_3$ respectively. Thus for a time $t<t^*$, $U$ is the
  sum of two kinks while for $t>t^*$, it is only one kink.  This illustrates
  soliton fusion.}
\end{figure}

\begin{figure}
\begin{center}
\includegraphics[height=2.5in]{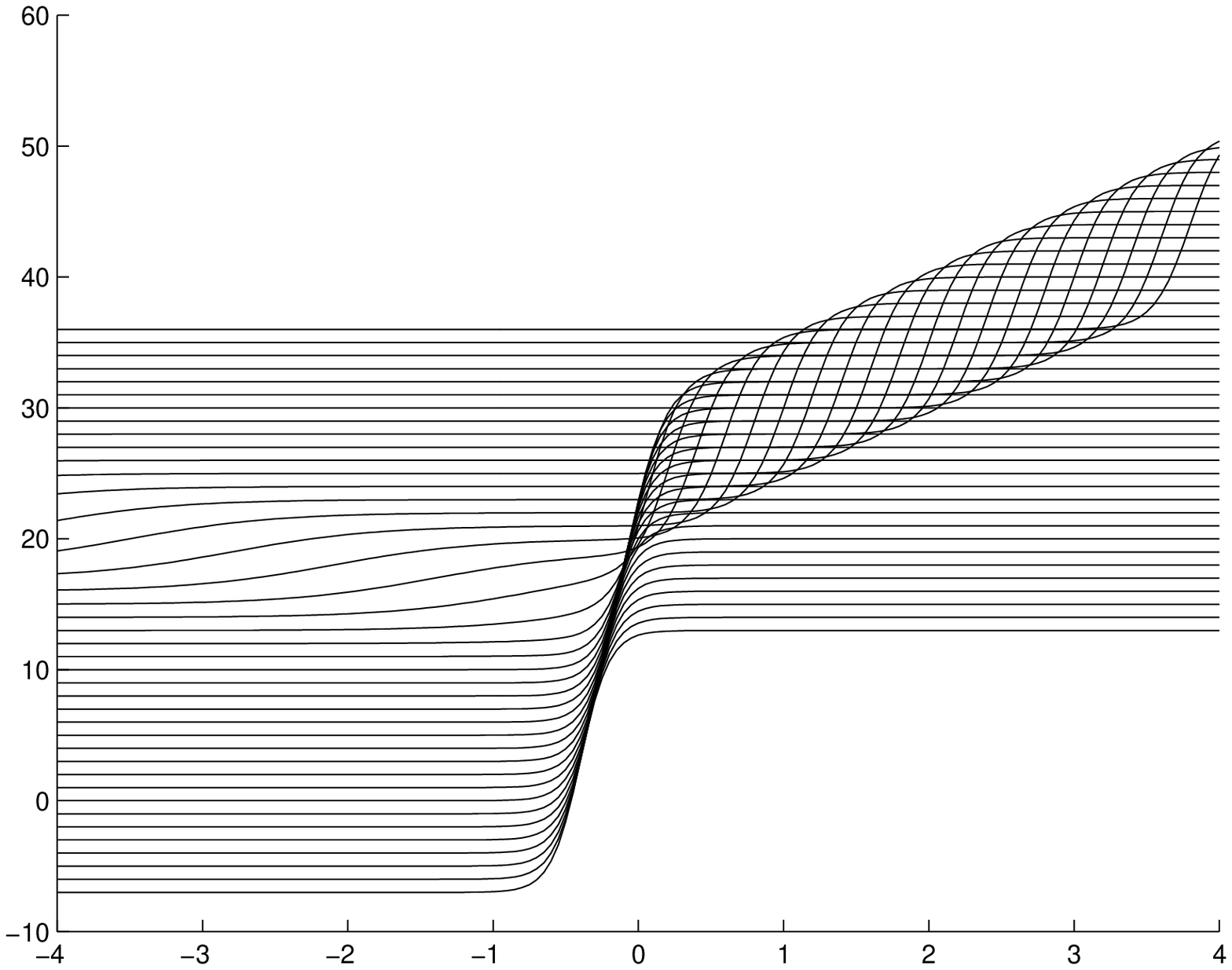}~\includegraphics[height=2.5in]
{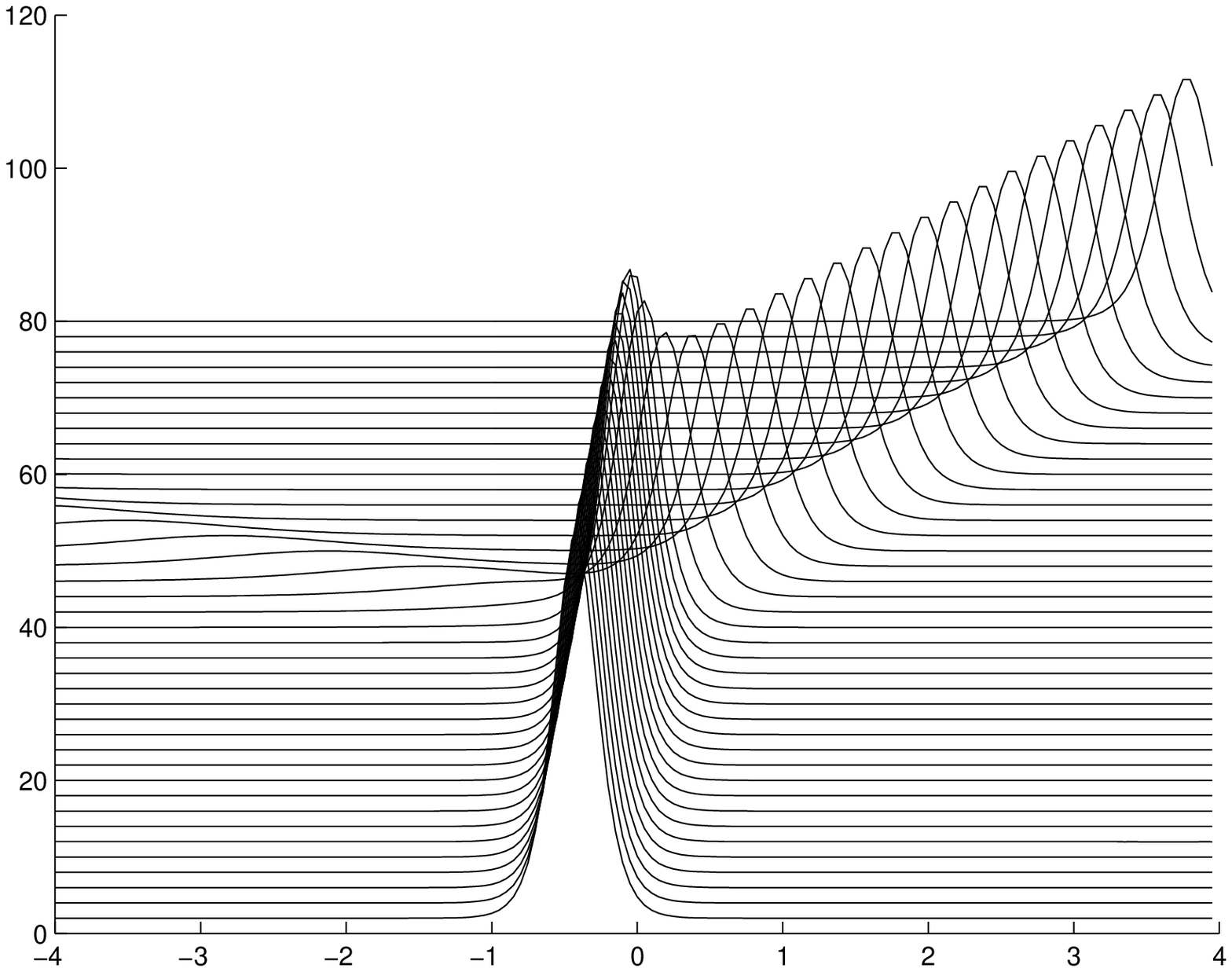}
\end{center}
\caption{\label{fig3cd}Genus-two solutions  $U$ and $W$ to the
  Boussinesq equation with two double roots.  Here  $a1=-2$,
  $a2=2$,$a_3=5$,$t=-.4 \dots .4$.  Both $\mu$ variables are
  initially chosen on the positive branch of the Riemann surface.
  Observe that each variable $U$ and $W$ undergoes soliton fission.}
\end{figure}

\begin{figure}
\begin{center}
\includegraphics[height=2.5in]{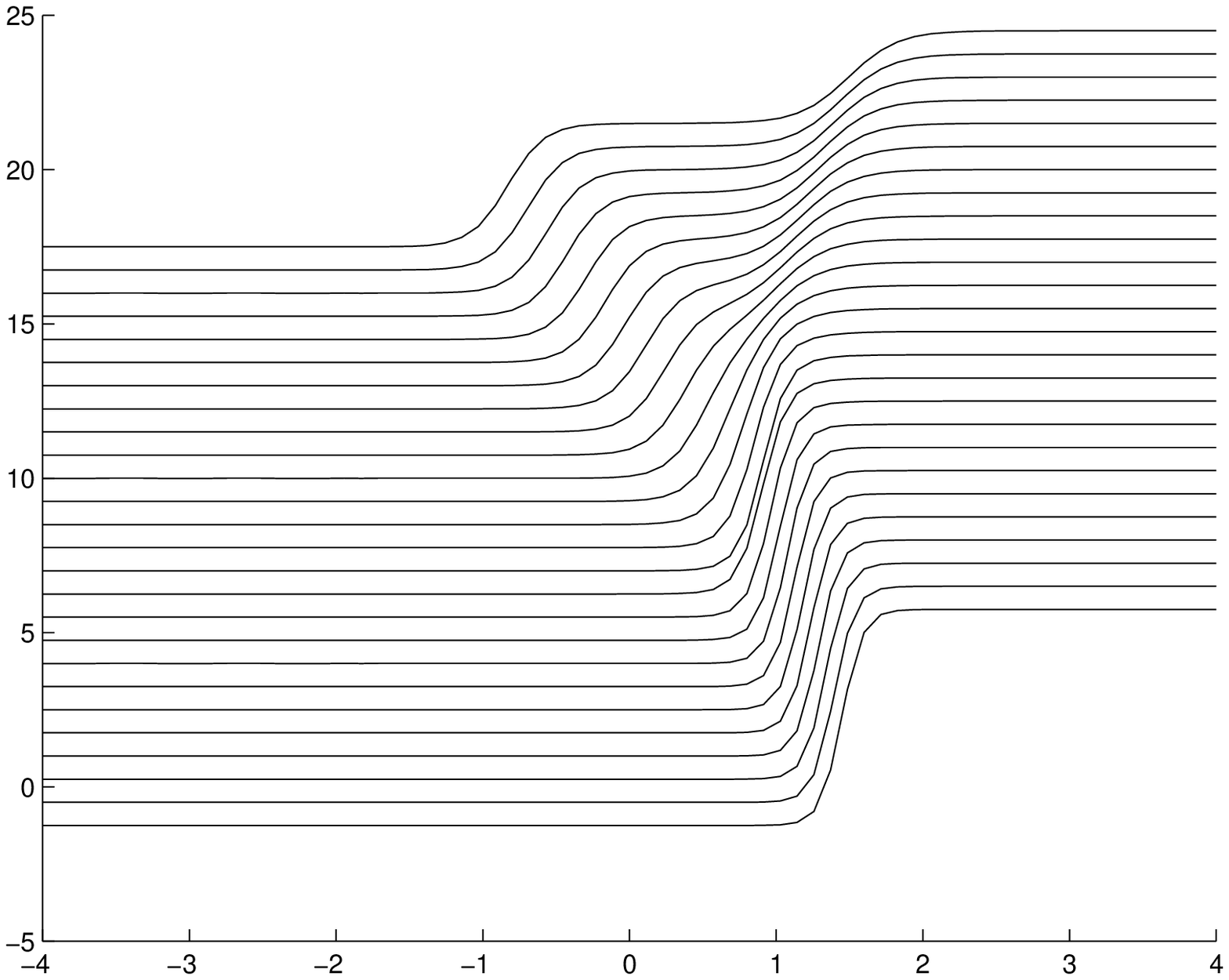}~\includegraphics[height=2.5in]
{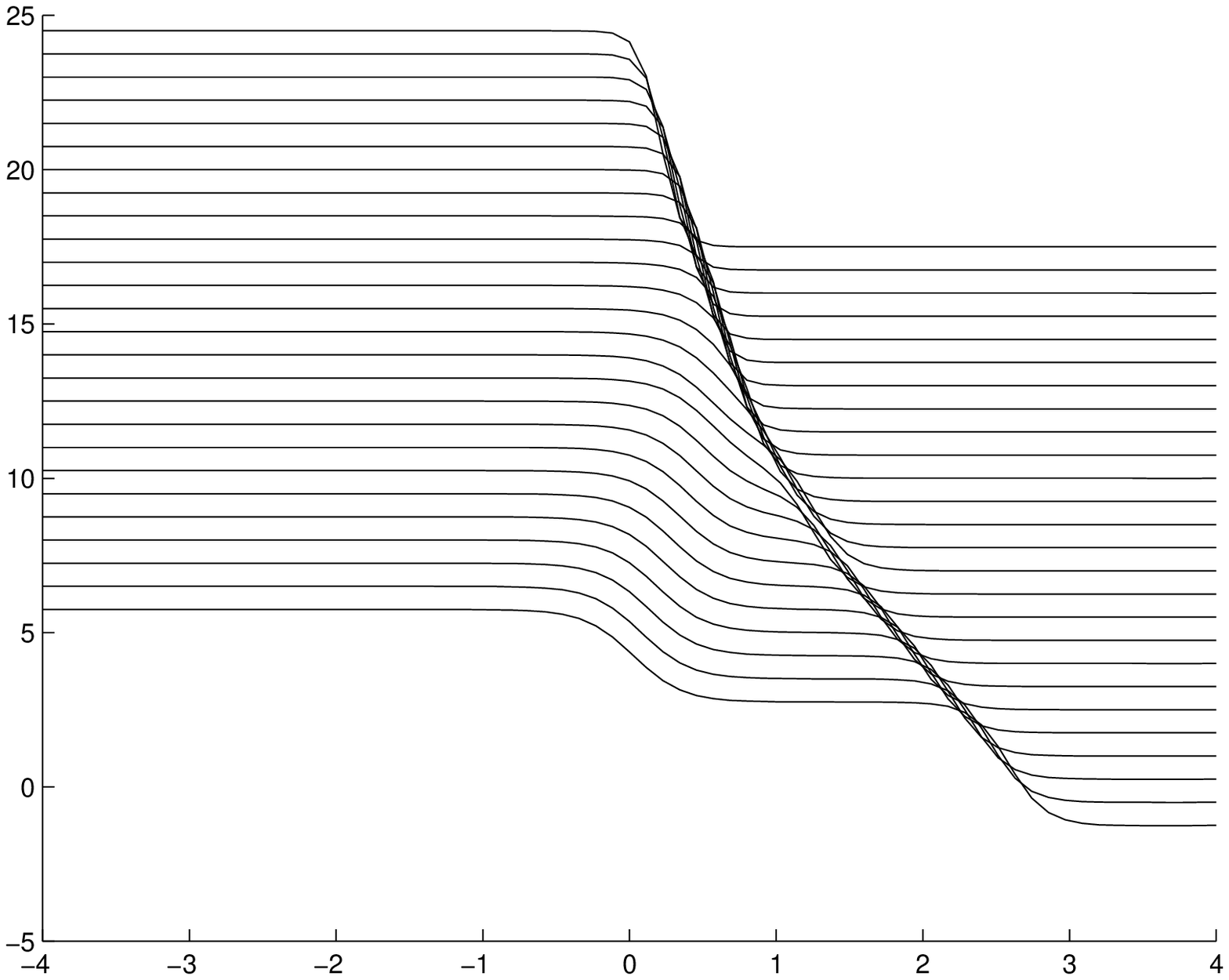}
\end{center}
\caption{\label{fig4ab}Genus-two $\mu_1$ and $\mu_2$ from
  (\ref{mu1cof}) and (\ref{mu2cof})
  illustrating when $\mu_1$ and $\mu_2$ are initially on different
  branches of the Riemann surface.  The parameters chosen for
  these graphs are $a1=-2$, $a2=0$,$a3=2$,$t=-2 \dots 2$.
  Notice how similar these graphs are to the fission/fusion of $U$
  in figure \ref{fig3cd}}
\end{figure}

\begin{figure}
\begin{center}
\includegraphics[height=2.5in]{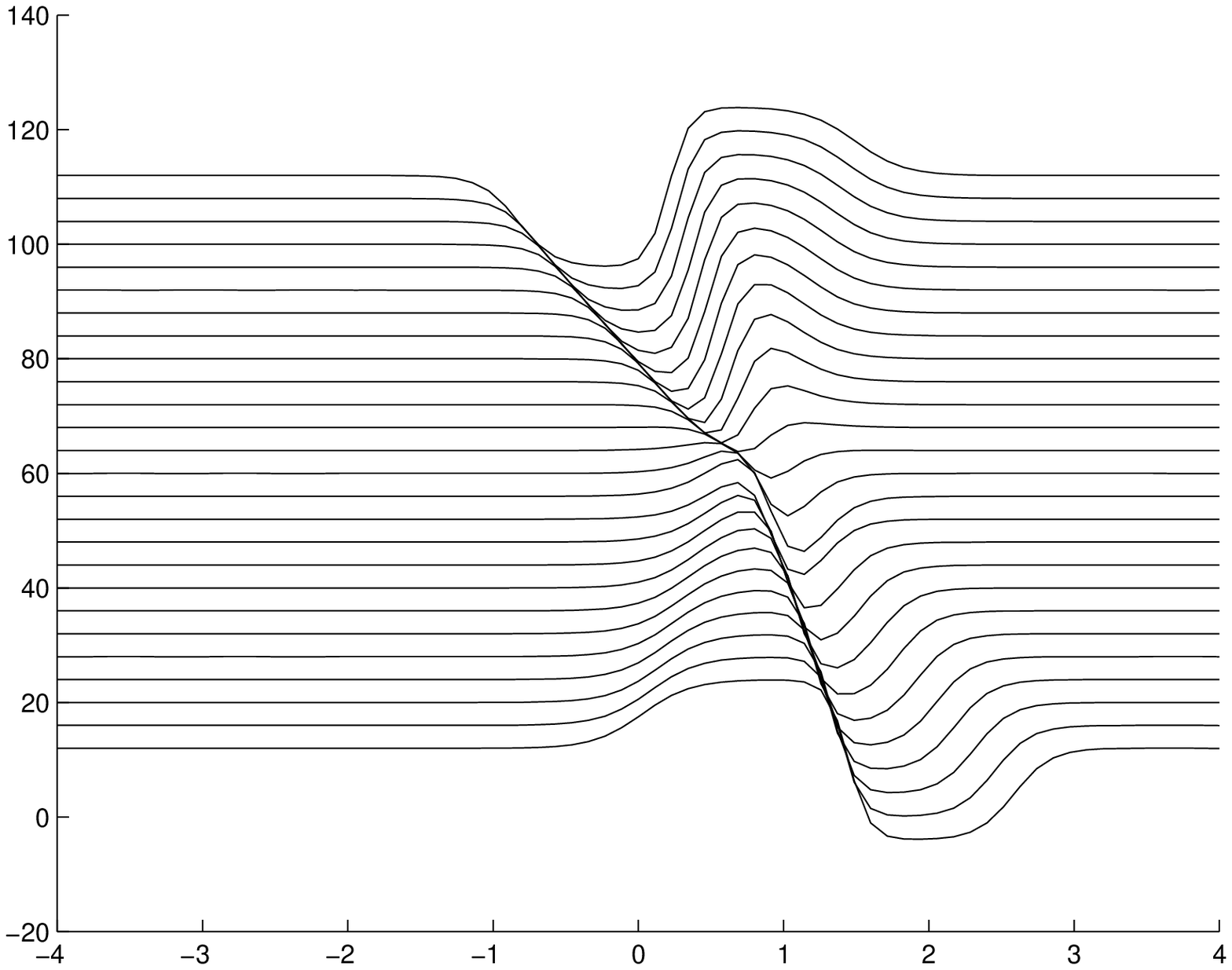}~\includegraphics[height=2.5in]
{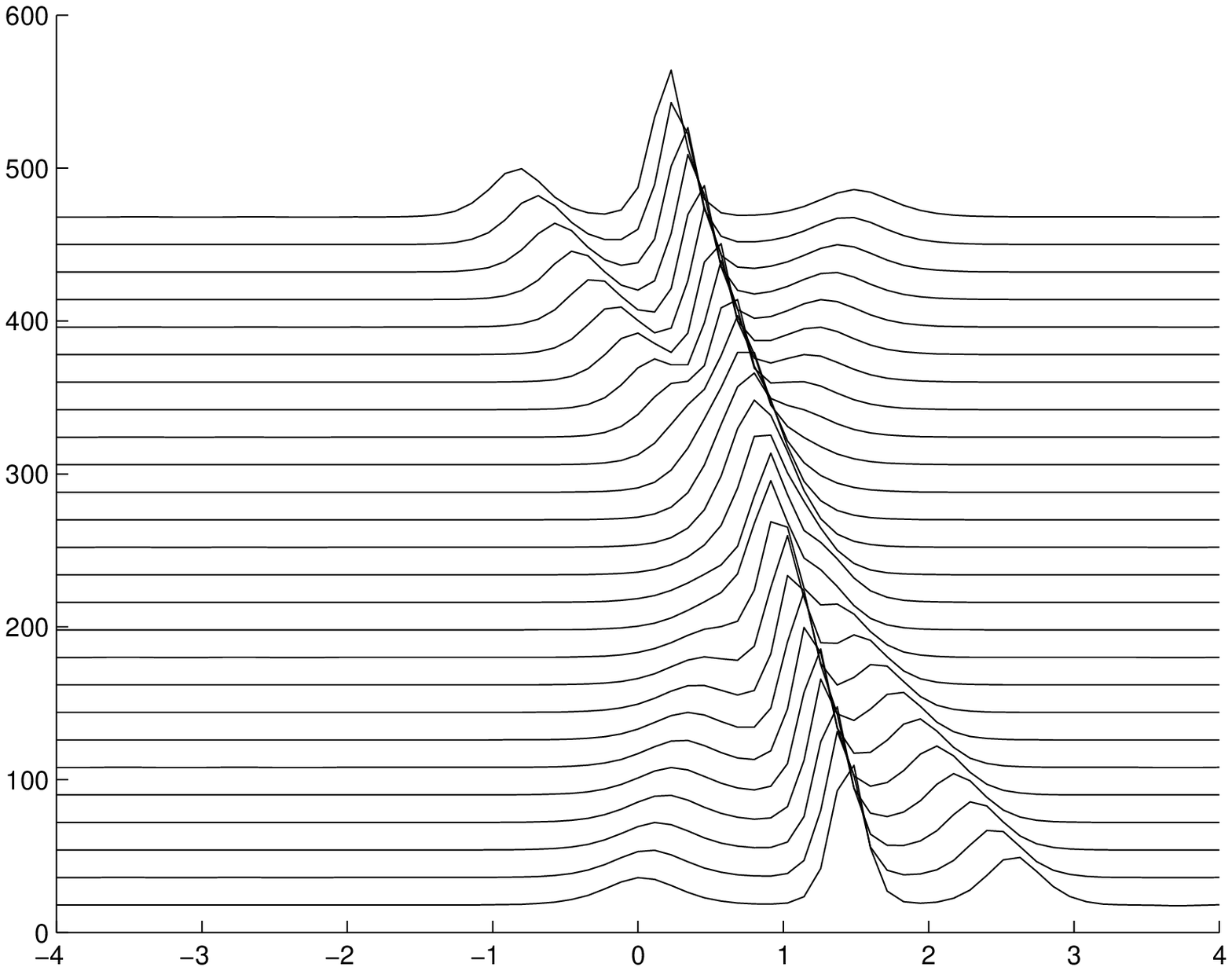}
\end{center}
\caption{\label{fig4cd}Genus-two $U$ and $W$ from the Boussinesq
  equation derived from (\ref{mu1cof}),(\ref{mu2cof}) and
  (\ref{coupledKdV1}), (\ref{coupledKdV2}).  Here
  $a1=-2$, $a2=2$,$a3=5$,$t=-.4 \dots .4$.  Here the $\mu$ variables
  from (\ref{mu1cof}) and (\ref{mu2cof}) are chosen to be on the
  opposite branches of the Riemann surface.   Notice the change of
  form in $U$ as kinks become antikinks and vice versa.}
\end{figure}

\begin{figure}
\begin{center}
\includegraphics[height=3in]{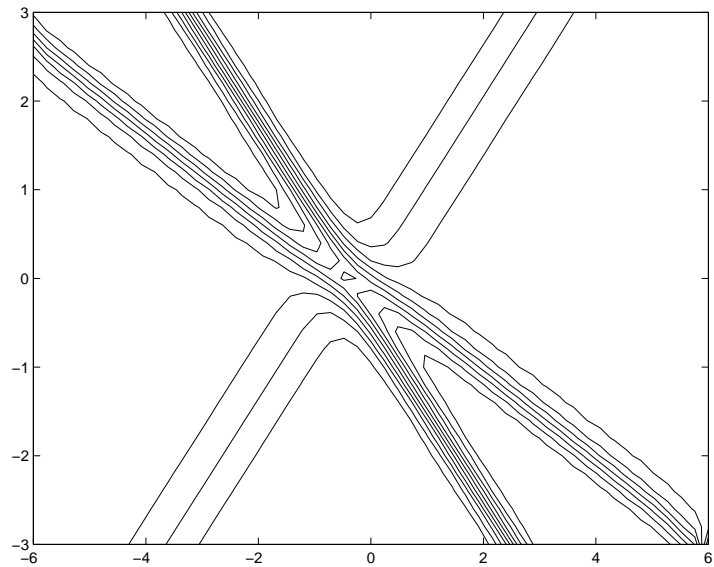} \end{center}
\caption{\label{fig:change}A contour plot of a genus-two solution
  $U$ of the Boussinesq equation for the parameters  $a1=-1$,
  $a2=1$,$a3=2$,$t=-3 \dots 3$.  This shows how the $\mu$
  variables combine to form different parts of the same kink.}
\end{figure}

\begin{figure}
\begin{center}
\includegraphics[height=3in]{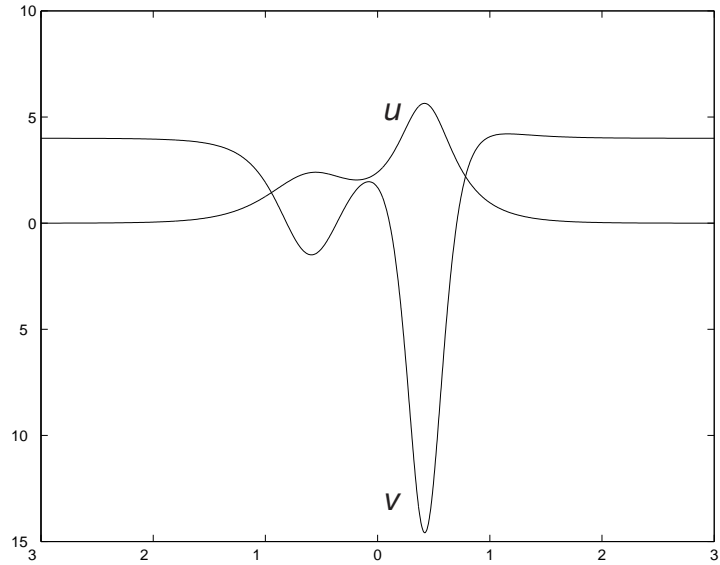} \end{center}
\caption{\label{fig:shg}A plot of $u$ and $v$, genus-two solutions
  derived from the $\mu$ variables in (\ref{an1}) and
  (\ref{an1part2}), except that $\kappa =-1$, using the trace
  formulas.  These can be transformed into the SHG equation
  using transformations in appendix A.  Here ($\kappa=-1$) for $a1=-1$,
  $a2=1$,$m_5=2$,$m_6=-2$,$t=0$, and the $\mu$ variables are
  initially chosen on opposite branches of the Riemann surface.}
\end{figure}

\begin{figure}
\caption{\label{fig:cdymkink}A genus-1 kink solution, $u(x,0)$,
for the coupled Dym
equation. The parameters used were
  two double roots $a1=1$, $a2=2$, and $t=0$}
\end{figure}
\begin{figure}
\begin{center}
\includegraphics[height=3in]{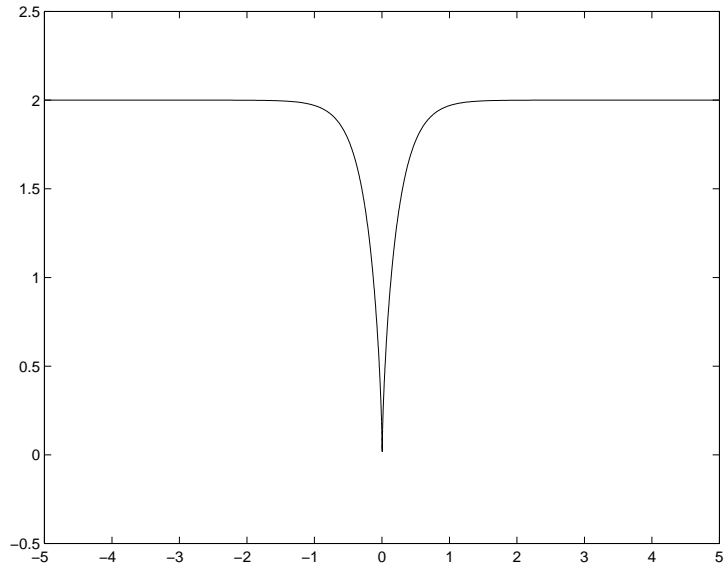} \end{center}
\caption{\label{fig:cdymcusp}A genus-1 cusp solution for the cDym equation.
 $u$ for two double roots $a1=-1$, $a2=1$, and $t=0$.}
\end{figure}

\begin{figure}
\begin{center}
\includegraphics[height=3in]{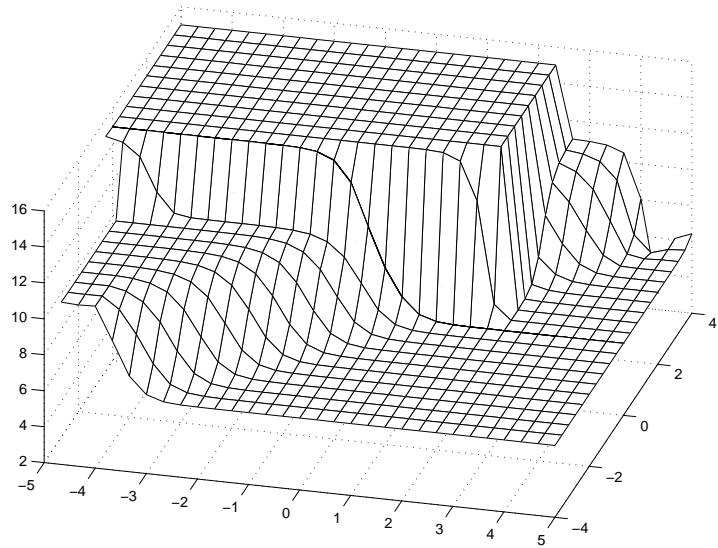} \end{center}
\caption{\label{fig:dymfission}Genus-two solution $u$ for the
  cDym equation with
  three double roots $a1=.1$, $a2=.5$,$a_3=1$,$t=-12 \dots 12$.
  Here the $\mu$ variables from which $u$ was derived were initially
  chosen to lie on the same branch of the Riemann surface.  In this
  situation soliton fission occurs as in the ckdv case. }
\end{figure}

\begin{figure}
\begin{center}
\includegraphics[height=3in]{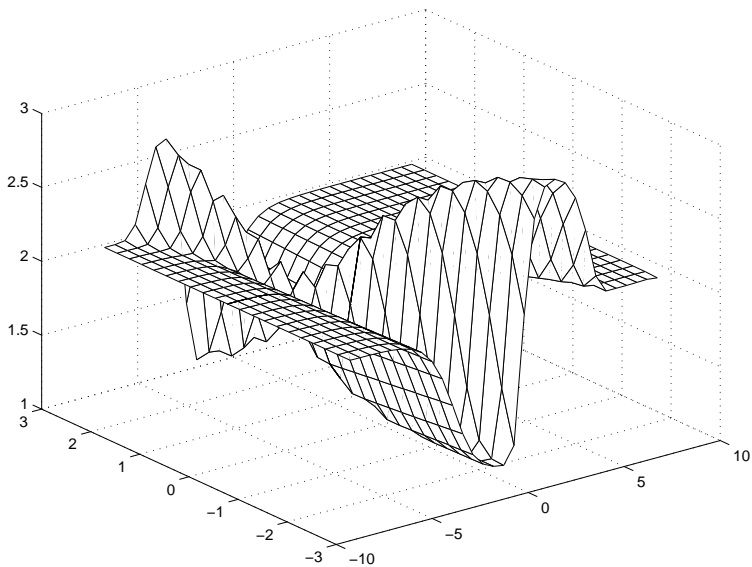} \end{center}
\caption{\label{fig:dymchange}Change of form for the Coupled Dym
  equation is seen for the genus-two solution $u$ with  three
  double roots $a1=1$, $a2=5$,$a_3=10$,$t=-.3 \dots .3$.  The
  $\mu$ variables from which $u$ was derived were initially chosen
  on opposite branches of the Riemann surface.  }
\end{figure}

\end{document}